\documentclass[runningheads,fleqn]{svmult}

\usepackage{subfigure}
\usepackage{epsfig}
\usepackage{psfrag}
\usepackage{rotating}
\usepackage{axodraw}

\usepackage{makeidx}   
\usepackage{graphicx}  
\usepackage{subeqnar}  
\usepackage{multicol}  
\usepackage{physmult}  
\makeindex             

\newcommand{\RE}{{\rm Re}}
\newcommand{\IM}{{\rm Im}}
\newcommand{\vcb}{|V_{cb}|}
\newcommand{\vtd}{|V_{td}|}
\newcommand{\vub}{|V_{ub}/V_{cb}|}
\newcommand{\vts}{|V_{ts}|}
\newcommand{\vus}{|V_{us}|}

\def\eps{\varepsilon}
\def\epe{\varepsilon'/\varepsilon}

\newcommand{\eqn}{\ref}
\newcommand{\mt}{m_{\rm t}}
\newcommand{\mtb}{\overline{m}_{\rm t}}

\newcommand{\mc}{m_{\rm c}}

\newcommand{\mw}{M_{\rm W}}

\newcommand{\gev}{\, {\rm GeV}}

\newcommand{\mev}{\, {\rm MeV}}
\newcommand{\bsi}{B_6^{(1/2)}}
\newcommand{\bei}{B_8^{(3/2)}}
\newcommand{\Lms}{\Lambda_{\overline{\rm MS}}}
\newcommand{\bea}{\begin{eqnarray}}
\newcommand{\eea}{\end{eqnarray}}
\newcommand{\bd}{\begin{displaymath}}
\newcommand{\ed}{\end{displaymath}}
\newcommand{\beq}{\begin{equation}}
\newcommand{\eeq}{\end{equation}}
\newcommand{\be}{\begin{equation}}
\newcommand{\ee}{\end{equation}}
\newcommand{\bi}{\begin{itemize}}
\newcommand{\ei}{\end{itemize}}
\newcommand{\ord}{{\cal O}}

\def\kpn{K^+\rightarrow\pi^+\nu\bar\nu}

\def\klpn{K_{\rm L}\rightarrow\pi^0\nu\bar\nu}

\begin{document}
%


\thispagestyle{empty}
\phantom{xxx}
\vskip1truecm
\begin{flushright}
 TUM-HEP-523/03 \\
 hep-ph/0307203 \\  
July 2003
\end{flushright}
\vskip1.8truecm
\centerline{\LARGE\bf CP Violation in B and K Decays: 2003}
   \vskip1truecm
\centerline{\Large\bf Andrzej J. Buras}
\bigskip
\centerline{\sl Technische Universit{\"a}t M{\"u}nchen}
\centerline{\sl Physik Department} 
\centerline{\sl D-85748 Garching, Germany}
\vskip1truecm
\centerline{\bf Abstract}
These lectures give a brief description of
CP violation in B and K meson decays with 
particular emphasize put on the determination of the CKM matrix.
The following topics will be discussed:
i) The CKM matrix, the unitarity triangle and general aspects of 
the theoretical framework, 
ii) Particle-antiparticle mixing and various types of CP violation,
iii) Standard analysis of the unitarity
triangle, 
iv) The ratio $\epe$, v) The most important strategies for the determination 
of the 
angles $\alpha$, $\beta$ and $\gamma$ from B decays, 
vi) Rare decays $K^+\to\pi^+\nu\bar\nu$ and $K_L\to\pi^0\nu\bar\nu$
and vii) Models with minimal flavour violation, in particular those with 
universal extra dimensions. 

\vskip1.6truecm

\centerline{\it  Lectures given at }
\centerline{\bf 41. Schladming School in Theoretical Physics}
\centerline{\it Schladming, February 22 -- 28, 2003}

\newpage
\setcounter{page}{0}

\title*{CP Violation in B and K Decays: 2003}
\toctitle{CP Violation in B and K Decays: 2003}
%
%
\titlerunning{CP Violation in B and K Decays:2003}
%
\author{Andrzej J. Buras}

\authorrunning{Andrzej J. Buras}
%
%

\maketitle              



\section{Introduction}
\subsection{Preface}
CP violation in B and K meson decays is not surprisingly one of the central 
topics in particle physics. Indeed, CP-violating and rare decays of 
K and B mesons are very sensitive to the flavour structure of the Standard 
Model (SM) and its extensions. In this context a very important role 
is played by the Cabibbo-Kobayashi-Maskawa (CKM) matrix \cite{CAB,KM}
that parametrizes the weak charged current interactions of quarks: 
its departure from the unit matrix is the origin of all flavour violating 
and CP-violating transitions in the SM and its simplest extensions.

One of the important questions still to be answered, is whether the 
CKM matrix is capable to describe with its the four parameters all weak 
decays that include in addition to tree level 
decays mediated by $W^\pm$-bosons, a vast number of one-loop induced flavour 
changing neutral current transitions involving gluons, photon, $W^\pm$, 
$Z^0$ and $H^0$. The latter transitions are responsible for rare decays and
CP-violating decays in the SM. This important role of the CKM matrix is
preserved in any extension of the SM even if more complicated extentions 
may contain new sources of flavour violation and CP violation.

The answer to this question is very challenging because the relevant rare 
and CP-violating decays have small branching ratios and are often very 
difficult to measure. Moreover, as hadrons are bound states of quarks and
antiquarks, the determination of the CKM parameters requires in many cases 
a quantitative control over QCD effects at long distances where the 
existing non-perturbative methods are not yet satisfactory. 

In spite of these difficulties, we strongly believe that this important 
question will be answered in this decade. This belief is based on an 
impressive 
progress in the experimental measurements in this field and on a 
similar progress made by theorists in  perturbative and 
to a lesser extend non-perturbative QCD calculations. The development of 
various strategies for the determination of the CKM parameters, 
that are essentially free from hadronic uncertainties, is also an important 
ingredient in this progress.
A recent account of these joined efforts by experimentalist and theorists is
given in \cite{CERNCKM}.

These lecture notes provide a rather non-technical description of the decays 
that are best suited for the determination of the CKM matrix. 
We will also briefly discuss 
the ratio $\epe$ that from the present
perspective is not suited for a precise determination of the CKM matrix 
but is interesting on its own.
There is unavoidably an overlap with our Les Houches  
\cite{AJBLH}, Lake Louise \cite{AJBLAKE}, Erice \cite{Erice} and 
Zacatecas \cite{MEX01} lectures and with the 
reviews \cite{BBL} and \cite{BF97}.
On the other hand new developments until the summer 2003 have been taken
into account, as far as the space allowed for it, and all numerical
results have been  updated. Moreover the discussion of the strategies for 
the determination of the angles $\alpha$, $\beta$ and $\gamma$ in the 
unitarity 
triangle goes far beyond our previous lectures. 

We hope that these lecture notes
will be helpful in following the new developments in this exciting field. 
In this respect
the recent books \cite{BULIND,Branco,Bigi}, the working group reports
\cite{CERNCKM,BABAR,LHCB,FERMILAB} and most recent reviews \cite{REV} are 
also strongly recommended.

\subsection{CKM Matrix and the Unitarity Triangle}

The unitary CKM matrix \cite{CAB,KM} connects  the {\it weak
eigenstates} $(d^\prime,s^\prime,b^\prime)$ and 
 the corresponding {\it mass eigenstates} $d,s,b$:
\begin{equation}\label{2.67}
\left(\begin{array}{c}
d^\prime \\ s^\prime \\ b^\prime
\end{array}\right)=
\left(\begin{array}{ccc}
V_{ud}&V_{us}&V_{ub}\\
V_{cd}&V_{cs}&V_{cb}\\
V_{td}&V_{ts}&V_{tb}
\end{array}\right)
\left(\begin{array}{c}
d \\ s \\ b
\end{array}\right)\equiv\hat V_{\rm CKM}\left(\begin{array}{c}
d \\ s \\ b
\end{array}\right).
\end{equation}

Many parametrizations of the CKM
matrix have been proposed in the literature. The classification of different 
parametrizations can be found in \cite{FX1}. While the so called 
standard parametrization \cite{CHAU} 
\begin{equation}\label{2.72}
\hat V_{\rm CKM}=
\left(\begin{array}{ccc}
c_{12}c_{13}&s_{12}c_{13}&s_{13}e^{-i\delta}\\ -s_{12}c_{23}
-c_{12}s_{23}s_{13}e^{i\delta}&c_{12}c_{23}-s_{12}s_{23}s_{13}e^{i\delta}&
s_{23}c_{13}\\ s_{12}s_{23}-c_{12}c_{23}s_{13}e^{i\delta}&-s_{23}c_{12}
-s_{12}c_{23}s_{13}e^{i\delta}&c_{23}c_{13}
\end{array}\right)\,,
\end{equation}
with
$c_{ij}=\cos\theta_{ij}$ and $s_{ij}=\sin\theta_{ij}$ 
($i,j=1,2,3$) and the complex phase $\delta$ necessary for {\rm CP} violation,
should be recommended \cite{PDG} 
for any numerical 
analysis, a generalization of the Wolfenstein parametrization \cite{WO} as 
presented in \cite{BLO} is more suitable for these lectures. 
On the one hand it is more transparent than the standard parametrization and 
on the other hand it  satisfies the unitarity 
of the CKM matrix to higher accuracy  than the original parametrization 
in \cite{WO}. 

To this end we make the following change of variables in
the standard parametrization (\ref{2.72}) 
\cite{BLO,schubert}
\begin{equation}\label{2.77} 
s_{12}=\lambda\,,
\qquad
s_{23}=A \lambda^2\,,
\qquad
s_{13} e^{-i\delta}=A \lambda^3 (\varrho-i \eta)
\end{equation}
where
\begin{equation}\label{2.76}
\lambda, \qquad A, \qquad \varrho, \qquad \eta \, 
\end{equation}
are the Wolfenstein parameters
with $\lambda\approx 0.22$ being an expansion parameter. We find then 
\be\label{f1}
V_{ud}=1-\frac{1}{2}\lambda^2-\frac{1}{8}\lambda^4, \qquad
V_{cs}= 1-\frac{1}{2}\lambda^2-\frac{1}{8}\lambda^4(1+4 A^2),
\ee
\be
V_{tb}=1-\frac{1}{2} A^2\lambda^4, \qquad
V_{cd}=-\lambda+\frac{1}{2} A^2\lambda^5 [1-2 (\varrho+i \eta)],
\ee
\be\label{VUS}
V_{us}=\lambda+\ord(\lambda^7),\qquad 
V_{ub}=A \lambda^3 (\varrho-i \eta), \qquad 
V_{cb}=A\lambda^2+\ord(\lambda^8),
\ee
\begin{equation}\label{2.83d}
 V_{ts}= -A\lambda^2+\frac{1}{2}A\lambda^4[1-2 (\varrho+i\eta)],
\qquad V_{td}=A\lambda^3(1-\bar\varrho-i\bar\eta)
\end{equation}
where terms 
$\ord(\lambda^6)$ and higher order terms have been neglected.
A non-vanishing $\eta$ is responsible for CP violation in the SM. It plays 
the role of $\delta$ in the standard parametrization.
Finally, the bared variables in (\ref{2.83d}) are given by
\cite{BLO}
\begin{equation}\label{2.88d}
\bar\varrho=\varrho (1-\frac{\lambda^2}{2}),
\qquad
\bar\eta=\eta (1-\frac{\lambda^2}{2}).
\end{equation}

Now, the unitarity of the CKM-matrix implies various relations between its
elements. In particular, we have
\begin{equation}\label{2.87h}
V_{ud}^{}V_{ub}^* + V_{cd}^{}V_{cb}^* + V_{td}^{}V_{tb}^* =0.
\end{equation}
The relation (\ref{2.87h})  can be
represented as a ``unitarity'' triangle in the complex 
$(\bar\varrho,\bar\eta)$ plane. 
One can construct additional five unitarity triangles \cite{Kayser} 
corresponding to other unitarity relations.

Noting that to an excellent accuracy $V_{cd}^{}V_{cb}^*$ is real with
$| V_{cd}^{}V_{cb}^*|=A\lambda^3+\ord(\lambda^7)$ and
rescaling all terms in (\ref{2.87h}) by $A \lambda^3$ 
we indeed find that the relation (\ref{2.87h}) can be represented 
as the triangle 
in the complex $(\bar\varrho,\bar\eta)$ plane 
as shown in fig.~\ref{fig:utriangle}. Let us collect useful formulae related 
to this triangle:

\begin{figure}[hbt]
\vspace{0.10in}
\centerline{
\epsfysize=2.0in
\epsffile{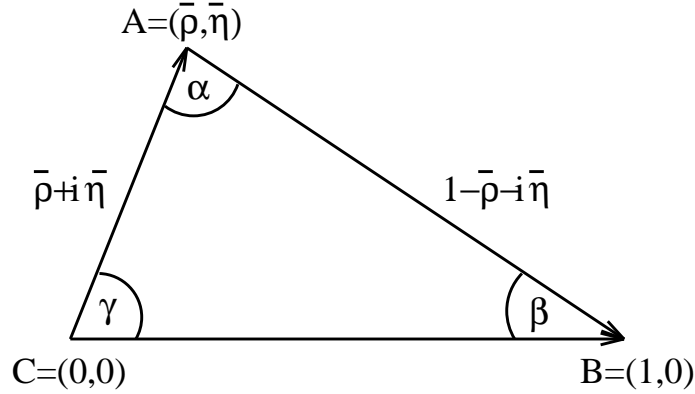}
}
\vspace{0.08in}
\caption{Unitarity Triangle.}\label{fig:utriangle}
\end{figure}

\bi
\item
We can express $\sin(2\phi_i$), $\phi_i=
\alpha, \beta, \gamma$, in terms of $(\bar\varrho,\bar\eta)$. In particular:
\begin{equation}\label{2.90}
\sin(2\beta)=\frac{2\bar\eta(1-\bar\varrho)}{(1-\bar\varrho)^2 + \bar\eta^2}.
\end{equation}
\item
The lengths $CA$ and $BA$ are given respectively by
\begin{equation}\label{2.94}
R_b \equiv \frac{| V_{ud}^{}V^*_{ub}|}{| V_{cd}^{}V^*_{cb}|}
= \sqrt{\bar\varrho^2 +\bar\eta^2}
= (1-\frac{\lambda^2}{2})\frac{1}{\lambda}
\left| \frac{V_{ub}}{V_{cb}} \right|,
\end{equation}
\begin{equation}\label{2.95}
R_t \equiv \frac{| V_{td}^{}V^*_{tb}|}{| V_{cd}^{}V^*_{cb}|} =
 \sqrt{(1-\bar\varrho)^2 +\bar\eta^2}
=\frac{1}{\lambda} \left| \frac{V_{td}}{V_{cb}} \right|.
\end{equation}
\item
The angles $\beta$ and $\gamma=\delta$ of the unitarity triangle 
are related
directly to the complex phases of the CKM-elements $V_{td}$ and
$V_{ub}$, respectively, through
\beq\label{e417}
V_{td}=|V_{td}|e^{-i\beta},\quad V_{ub}=|V_{ub}|e^{-i\gamma}.
\eeq
\item
The unitarity relation (\ref{2.87h}) can be rewritten as
\be\label{RbRt}
R_b e^{i\gamma} +R_t e^{-i\beta}=1~.
\ee
\item
The angle $\alpha$ can be obtained through the relation
\beq\label{e419}
\alpha+\beta+\gamma=180^\circ~.
\eeq
\ei

Formula (\ref{RbRt}) shows transparently that the knowledge of
$(R_t,\beta)$ allows to determine $(R_b,\gamma)$ through 
\be\label{VUBG}
R_b=\sqrt{1+R_t^2-2 R_t\cos\beta},\qquad
\cot\gamma=\frac{1-R_t\cos\beta}{R_t\sin\beta}.
\ee
Similarly, $(R_t,\beta)$ can be expressed through $(R_b,\gamma)$:
\be\label{VTDG}
R_t=\sqrt{1+R_b^2-2 R_b\cos\gamma},\qquad
\cot\beta=\frac{1-R_b\cos\gamma}{R_b\sin\gamma}.
\ee
These relations are remarkable. They imply that the knowledge 
of the coupling $V_{td}$ between $t$ and $d$ quarks allows to deduce the 
strength of the corresponding coupling $V_{ub}$ between $u$ and $b$ quarks 
and vice versa.

The triangle depicted in fig. \ref{fig:utriangle}, $|V_{us}|$ 
and $\vcb$ give the full description of the CKM matrix. 
Looking at the expressions for $R_b$ and $R_t$, we observe that within
the SM the measurements of four CP
{\it conserving } decays sensitive to $|V_{us}|$, $|V_{ub}|$,   
$|V_{cb}|$ and $|V_{td}|$ can tell us whether CP violation
($\bar\eta \not= 0$ or $\gamma \not=0,\pi$) is predicted in the SM. 
This fact is often used to determine
the angles of the unitarity triangle without the study of CP-violating
quantities.

\subsection{The Special Role of \boldmath{$|V_{us}|$}, \boldmath{$|V_{ub}|$}
and \boldmath{$|V_{cb}|$}}
What do we know about the CKM matrix and the unitarity triangle on the
basis of {\it tree level} decays? 
Here the semi-leptonic K and B decays play the decisive role. 
The present situation can be summarized by \cite{CERNCKM} 
\begin{equation}\label{vcb}
|V_{us}| = \lambda =  0.2240 \pm 0.0036\,
\quad\quad
\vcb=(41.5\pm0.8)\cdot 10^{-3},
\end{equation}
\begin{equation}\label{v13}
\frac{|V_{ub}|}{\vcb}=0.086\pm0.008, \quad\quad
|V_{ub}|=(3.57\pm0.31)\cdot 10^{-3}.
\end{equation}
implying
\be
 A=0.83\pm0.02,\qquad R_b=0.37\pm 0.04~.
\ee
There is an impressive work done by theorists and experimentalists hidden
behind these numbers. We refer to \cite{CERNCKM} for details.
See also \cite{PDG}.

\begin{figure}[hbt]
\vspace{-0.10in}
\centerline{
\epsfysize=2.0in
\epsffile{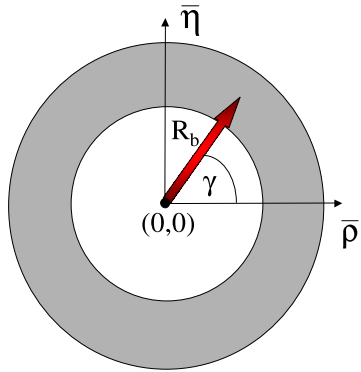}
}
\vspace{0.02in}
\caption[]{``Unitarity Clock".
\label{L:2}}
\end{figure}

The information given above tells us only that the apex $A$ of the unitarity 
triangle lies in the band shown in fig.~\ref{L:2}. 
While this information appears at first sight to be rather limited, 
it is very important for the following reason. As $|V_{us}|$, $\vcb$, 
 $|V_{ub}|$ and consequently $R_b$ are determined here from tree level 
decays, their
values given above are to an excellent accuracy independent of any 
new physics contributions. They are universal fundamental 
constants valid in any extention of the SM. Therefore their precise 
determinations are of utmost importance. 
In order to answer the question where
the apex $A$ lies on the ``unitarity clock'' in fig.~\ref{L:2} we have to 
look at other decays. Most promising in this respect are the so-called 
``loop induced'' decays and transitions and CP-violating B decays. 
These decays are sensitive to the angles $\beta$ and $\gamma$ as well as 
to the length $R_t$ and measuring only one of these three quantities allows to 
find the unitarity triangle provided the universal $R_b$ is known.

Of course any pair among $(R_t,\beta,\gamma)$ is sufficient to construct 
the UT without any knowledge of $R_b$. Yet the special role of $R_b$ among
these variables lies in its universality whereas the other three variables 
are generally sensitive functions of possible new physics contributions.
This means that assuming three generation unitarity 
of the CKM matrix and that the SM is a part of a bigger 
theory, the apex of the unitarity triangle has to be eventually placed
on the unitarity clock with the radius $R_b$ obtained from tree level decays.
That is even 
if using SM expressions for loop induced processes, $(\bar\varrho,\bar\eta)$
would be found outside the unitarity clock, the corresponding expressions 
of the grander theory must include appropriate new contributions so that 
the apex of the unitarity triangle is shifted back to the band in  
fig.~\ref{L:2}. In the case of CP asymmetries this could be achieved by 
realizing that the measured angles $\alpha$, $\beta$ and $\gamma$ are not
the true angles of the unitarity triangle but sums of the true angles and 
new complex phases present in extentions of the SM. The better $R_b$ is known,
the thiner the band in fig.~\ref{L:2} will be, 
selecting in this manner efficiently the correct theory. On the other hand 
as the 
the branching ratios for rare and CP-violating decays depend sensitively
on the parameter $A$, the precise knowledge of $\vcb$ is also very important.

\subsection{Grand Picture}
\begin{figure}[hbt]
  \vspace{0.10in} \centerline{
\begin{turn}{-90}
  \mbox{\epsfig{file=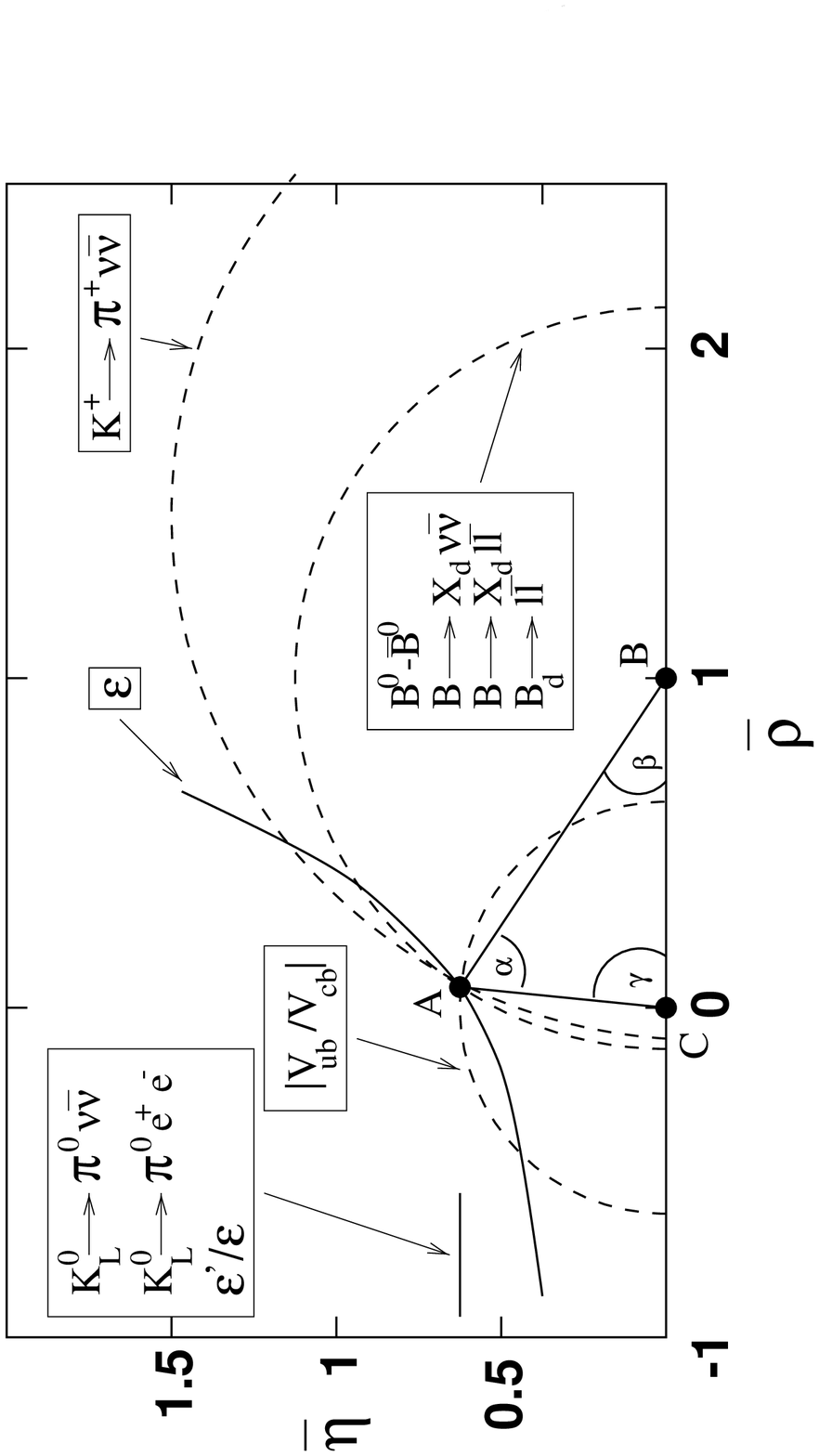,width=0.4\linewidth}}
\end{turn}
} \vspace{0.08in}
\caption[]{
  The ideal Unitarity Triangle.
\label{fig:2011}}
\end{figure}

The apex $(\bar\varrho,\bar\eta)$ of the UT can be efficiently hunted by 
means of 
rare and CP violating transitions as shown in fig.~\ref{fig:2011}. Moreover 
the angles 
of this triangle can be measured in CP asymmetries in B-decays and 
using other strategies. This picture could describe in principle the 
reality in the year 2012, my retirement year, if the SM is the whole story. 
On the other hand in the 
presence of significant new physics contributions, the use of the SM 
expressions for rare and CP violating transitions in question, combined 
with future precise measurements, may result in curves which do not cross 
each other at a single point in the $(\bar\varrho,\bar\eta)$ plane. 
This would be truly exciting and most of us hope that this will turn out 
to be the case. In order to be able to draw such thin curves as in 
fig.~\ref{fig:2011}, not only experiments but also the theory has to be 
under control. 

\subsection{Theoretical Framework}
The present framework for weak decays is based on the operator product 
expansion (OPE) that allows to separate short and long distance 
contributions to weak amplitudes and on the renormalization group (RG) 
methods that allow to sum large logarithms $\log \mu_{SD}/\mu_{LD}$  to 
all orders in perturbation theory. The full exposition of these methods 
can be found in \cite{AJBLH,BBL}. 

The OPE allows to write
the effective weak Hamiltonian simply as follows
\be\label{b1}
{\cal H}_{eff}=\frac{G_F}{\sqrt{2}}\sum_i V^i_{\rm CKM}C_i(\mu)Q_i~.
\ee
Here $G_F$ is the Fermi constant and $Q_i$ are the relevant local
operators which govern the decays in question. 
They are built out of quark and lepton fields.
The Cabibbo-Kobayashi-Maskawa
factors $V^i_{\rm CKM}$ \cite{CAB,KM} 
and the Wilson coefficients $C_i(\mu)$ describe the 
strength with which a given operator enters the Hamiltonian.
The latter coefficients can be considered as scale dependent
``couplings'' related to ``vertices'' $Q_i$ and as discussed below
can be calculated using perturbative methods as long as $\mu$ is
not too small.

An amplitude for a decay of a given meson 
$M= K, B,..$ into a final state $F=\pi\nu\bar\nu,~\pi\pi,~DK$ is then
simply given by
\be\label{amp5}
A(M\to F)=\langle F|{\cal H}_{eff}|M\rangle
=\frac{G_F}{\sqrt{2}}\sum_i V^i_{CKM}C_i(\mu)\langle F|Q_i(\mu)|M\rangle,
\ee
where $\langle F|Q_i(\mu)|M\rangle$ 
are the matrix elements of $Q_i$ between M and F, evaluated at the
renormalization scale $\mu$. 

The essential virtue of OPE is this one. It allows to separate the problem
of calculating the amplitude
$A(M\to F)$ into two distinct parts: the {\it short distance}
(perturbative) calculation of the coefficients $C_i(\mu)$ and 
the {\it long-distance} (generally non-perturbative) calculation of 
the matrix elements $\langle Q_i(\mu)\rangle$. The scale $\mu$
separates, roughly speaking, the physics contributions into short
distance contributions contained in $C_i(\mu)$ and the long distance 
contributions
contained in $\langle Q_i(\mu)\rangle$. 
Thus $C_i$ include the top quark contributions and
contributions from other heavy particles such as W-, Z-bosons and charged
Higgs particles or supersymmetric particles in the supersymmetric extensions
of the SM. 
Consequently $C_i(\mu)$ depend generally 
on $m_t$ and also on the masses of new particles if extensions of the 
SM are considered. This dependence can be found by evaluating 
so-called {\it box} and {\it penguin} diagrams with full W-, Z-, top- and 
new particles exchanges and {\it properly} including short distance QCD 
effects. The latter govern the $\mu$-dependence of $C_i(\mu)$.

The value of $\mu$ can be chosen arbitrarily but the final result
must be $\mu$-independent.
Therefore 
the $\mu$-dependence of $C_i(\mu)$ has to cancel the 
$\mu$-dependence of $\langle Q_i(\mu)\rangle$. 
The same comments apply to the renormalization scheme dependence of 
$C_i(\mu)$ and $\langle Q_i(\mu)\rangle$.

Now due to the fact that for low energy processes the appropriate scale 
 $\mu\ll  M_{W,Z},~ m_t$, large logarithms 
$\ln\mw/\mu$ compensate in the evaluation of
$C_i(\mu)$ the smallness of the QCD coupling constant $\alpha_s$ and 
terms $\alpha^n_s (\ln\mw/\mu)^n$, $\alpha^n_s (\ln\mw/\mu)^{n-1}$ 
etc. have to be resummed to all orders in $\alpha_s$ before a reliable 
result for $C_i$ can be obtained.
This can be done very efficiently by means of the renormalization group
methods. 
The resulting {\it renormalization group improved} perturbative
expansion for $C_i(\mu)$ in terms of the effective coupling constant 
$\alpha_s(\mu)$ does not involve large logarithms and is more reliable.
The related technical issues are discussed in detail in \cite{AJBLH}
and \cite{BBL}. It should be emphasized that by 2003 the next-to-leading 
(NLO) QCD and QED corrections to all relevant weak decay processes 
in the SM are known.

Clearly, in order to calculate the amplitude $A(M\to F)$ the matrix 
elements $\langle Q_i(\mu)\rangle$ have to be evaluated. 
Since they involve long distance contributions one is forced in
this case to use non-perturbative methods such as lattice calculations, the
1/N expansion (N is the number of colours), QCD sum rules, hadronic sum rules
and chiral perturbation theory. In the case of B-meson decays,
the {\it Heavy Quark Effective Theory} (HQET) and {\it Heavy Quark
Expansions} (HQE) also turn out to be useful tools.
However, all these non-perturbative methods have some limitations.
Consequently the dominant theoretical uncertainties in the decay amplitudes
reside in the matrix elements $\langle Q_i(\mu)\rangle$ and non-perturbative 
parameters present in HQET and HQE.
These issues are reviewed in \cite{CERNCKM}.

The fact that in many cases the matrix elements $\langle Q_i(\mu)\rangle$
 cannot be reliably
calculated at present, is very unfortunate. The main goals of the
experimental studies of weak decays is the determination of the CKM factors 
$V_{\rm CKM}$
and the search for the physics beyond the SM. Without a reliable
estimate of $\langle Q_i(\mu)\rangle$ these goals cannot be achieved unless 
these matrix elements can be determined experimentally or removed from the 
final measurable quantities
by taking suitable ratios and combinations of decay amplitudes or branching
ratios. 
We will encounter many examples in these lectures.
Flavour symmetries like $SU(2)_{\rm F}$ and 
$SU(3)_{\rm F}$ relating various
matrix elements can be useful in this respect, provided flavour
symmetry breaking effects can be reliably calculated. 
A recent progress in the calculation of $\langle Q_i(\mu)\rangle$ relevant
for non-leptonic B decays can be very helpful here as discussed in Section 5.

After these general remarks let us
be more specific about the structure of (\ref{amp5}) by considering the 
simplest class of models in which all flavour violating and CP-violating 
transition are governed by the CKM matrix and the only relevant local 
operators are the ones that are relevant in the SM.
We will call this scenario ``Minimal Flavour 
Violation" (MFV) \cite{UUT} being aware of the fact that for some authors 
MFV means a more general framework in which also new operators can give 
significant contributions. 
See for instance the recent discussions in 
\cite{BOEWKRUR,AMGIISST}.
In the MFV models, as defined in \cite{UUT},  the formula  (\ref{amp5})
can be written as follows
\be\label{mmaster}
{\rm A(Decay)}= \sum_i B_i \eta^i_{\rm QCD}V^i_{\rm CKM} F^i, \qquad 
F^i= F^i_{\rm SM}+F^i_{\rm New} 
\ee 
with $F^i_{\rm SM}$ and $F^i_{\rm New}$ being real. 

Here the non-perturbative parameters $B_i$ represent the matrix elements of 
local 
operators present in the SM. For instance in the case of 
$K^0-\bar K^0$ mixing, the matrix element of the operator
$\bar s \gamma_\mu(1-\gamma_5) d \otimes \bar s \gamma^\mu(1-\gamma_5) d $
is represented by the parameter $\hat B_K$.
There are other non-perturbative parameters in the SM that represent 
matrix elements of operators $Q_i$ with different colour and Dirac 
structures. The objects $\eta^i_{\rm QCD}$ are the QCD factors resulting 
from RG-analysis of the corresponding operators and $F^i_{\rm SM}$ stand for 
the so-called Inami-Lim functions \cite{IL} that result from the calculations 
of various
box and penguin diagrams. They depend on the top-quark mass. 
$V^i_{\rm CKM}$ are 
the CKM-factors we want to determine. 

The important point is that in all MFV models $B_i$ and 
$\eta^i_{\rm QCD}$ are the same as in the SM and the only place where 
the new physics enters are the new short distance functions $F^i_{\rm New}$ 
that depend on the new 
parameters in the extensions of the SM like the masses of charginos, 
squarks, charged Higgs particles and $\tan\beta=v_2/v_1$ in the MSSM. 
These new 
particles enter the new box and penguin diagrams.
Strictly speaking at the NLO level the QCD corrections to the new diagrams 
at scales larger than $\ord(\mw)$ may differ from the corresponding 
corrections in the SM but this, generally small, difference can be 
absorbed into $F^i_{\rm New}$ so that $\eta^i_{\rm QCD}$ are 
the QCD corrections calculated in the SM. Indeed, the QCD corrections at 
scales lower than $\ord(\mw)$ are related to the renormalization of the 
local operators that are common to all models in this class.

In more complicated 
extensions of the SM new operators (Dirac structures) that are either 
absent or very strongly suppressed in the SM, can become important.
Moreover new sources of flavour and CP violation beyond the CKM matrix, 
including new complex phases, could be present. A general master formula
describing such contributions is given in \cite{Pisa}.

Finally, let me give some arguments why our definition of MFV models is
phenomenologically useful. With a simple formula like (\ref{mmaster})
it is possible to derive a number of relations that are independent of
the parameters specific to a given MFV models. Consequently, any 
violation of these relations will signal the presence  
of new local operators and/or new complex phases that are necessary 
to describe the data. We will return to this point in Section 7.

\section{Particle-Antiparticle Mixing and Various Types\\ of CP
Violation}
        \label{sec:epsBBUT}
\subsection{Preliminaries}
Let us next discuss the formalism of particle--antiparticle mixing
and CP violation. Much more elaborate discussion can be found in
two books \cite{Branco,Bigi}. We will concentrate here on
$K^0-\bar K^0$ mixing, $B_{d,s}^0-\bar B^0_{d,s}$ mixings and
CP violation in K-meson and B-meson decays. 
Due to GIM mechanism \cite{GIM}
 the phenomena discussed in this section
appear first at the one--loop level and as such they are
sensitive measures of the top quark couplings $V_{ti}(i=d,s,b)$ and 
in particular of the phase $\delta=\gamma$.
They allow then to construct the unitarity triangle as explicitly
demonstrated in Section 4.

\begin{figure}[hbt]
\vspace{0.10in}
\centerline{
\epsfysize=1.5in
\epsffile{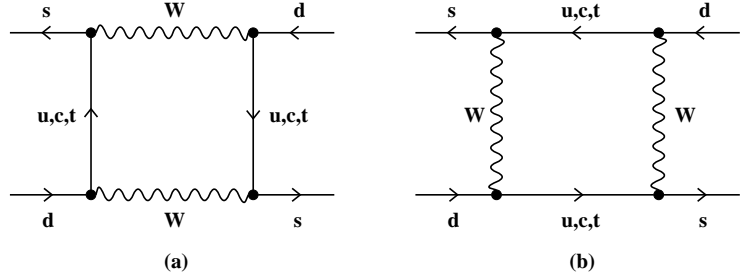}
}
\vspace{0.08in}
\caption[]{Box diagrams contributing to $K^0-\bar K^0$ mixing
in the SM.
\label{L:9}}
\end{figure}

\subsection{Express Review of $K^0-\bar K^0$ Mixing}
$K^0=(\bar s d)$ and $\bar K^0=(s\bar d)$ are flavour eigenstates which 
in the SM
may mix via weak interactions through the box diagrams in fig.
\ref{L:9}.
We will choose the phase conventions so that 
\be
CP|K^0\rangle=-|\bar K^0\rangle, \qquad   CP|\bar K^0\rangle=-|K^0\rangle.
\ee

In the absence of mixing the time evolution of $|K^0(t)\rangle$ is
given by
\be
|K^0(t)\rangle=|K^0(0)\rangle \exp(-iHt)~, 
\qquad H=M-i\frac{\Gamma}{2}~,
\ee
where $M$ is the mass and $\Gamma$ the width of $K^0$. Similar formula
exists for $\bar K^0$.

On the other hand, in the presence of flavour mixing the time evolution 
of the $K^0-\bar K^0$ system is described by
\be\label{SCH}
i\frac{d\psi(t)}{dt}=\hat H \psi(t) \qquad  
\psi(t)=
\left(\begin{array}{c}
|K^0(t)\rangle\\
|\bar K^0(t)\rangle
\end{array}\right)
\ee
where
\be
\hat H=\hat M-i\frac{\hat\Gamma}{2}
= \left(\begin{array}{cc} 
M_{11}-i\frac{\Gamma_{11}}{2} & M_{12}-i\frac{\Gamma_{12}}{2} \\
M_{21}-i\frac{\Gamma_{21}}{2}  & M_{22}-i\frac{\Gamma_{22}}{2}
    \end{array}\right)
\ee
with $\hat M$ and $\hat\Gamma$ being hermitian matrices having positive
(real) eigenvalues in analogy with $M$ and $\Gamma$. $M_{ij}$ and
$\Gamma_{ij}$ are the transition matrix elements from virtual and physical
intermediate states respectively.
Using
\be
M_{21}=M^*_{12}~, \qquad 
\Gamma_{21}=\Gamma_{12}^*~,\quad\quad {\rm (hermiticity)}
\ee
\be
M_{11}=M_{22}\equiv M~, \qquad \Gamma_{11}=\Gamma_{22}\equiv\Gamma~,
\quad {\rm (CPT)}
\ee
we have
\be\label{MM12}
\hat H=
 \left(\begin{array}{cc} 
M-i\frac{\Gamma}{2} & M_{12}-i\frac{\Gamma_{12}}{2} \\
M^*_{12}-i\frac{\Gamma^*_{12}}{2}  & M-i\frac{\Gamma}{2}
    \end{array}\right)~.
\ee

Diagonalizing (\ref{SCH}) we find:

{\bf Eigenstates:}
\be\label{KLS}
K_{L,S}=\frac{(1+\bar\varepsilon)K^0\pm (1-\bar\varepsilon)\bar K^0}
        {\sqrt{2(1+\mid\bar\varepsilon\mid^2)}}
\ee
where $\bar\varepsilon$ is a small complex parameter given by
\be\label{bare3}
\frac{1-\bar\varepsilon}{1+\bar\varepsilon}=
\sqrt{\frac{M^*_{12}-i\frac{1}{2}\Gamma^*_{12}}
{M_{12}-i\frac{1}{2}\Gamma_{12}}}=
\frac{2 M^*_{12}-i\Gamma^*_{12}}{\Delta M-i\frac{1}{2}\Delta\Gamma}
\equiv r\exp(i\kappa)~.
\ee
with $\Delta\Gamma$ and $\Delta M$ given below.

{\bf Eigenvalues:}
\be
M_{L,S}=M\pm \RE Q  \qquad \Gamma_{L,S}=\Gamma\mp 2 \IM Q
\ee
where
\be
Q=\sqrt{(M_{12}-i\frac{1}{2}\Gamma_{12})(M^*_{12}-i\frac{1}{2}\Gamma^*_{12})}.
\ee
Consequently we have
\be\label{deltak}
\Delta M= M_L-M_S = 2\RE Q~,
\quad\quad
\Delta\Gamma=\Gamma_L-\Gamma_S=-4 \IM Q.
\ee

It should be noted that the mass eigenstates $K_S$ and $K_L$ differ from 
CP eigenstates
\begin{equation}
K_1={1\over{\sqrt 2}}(K^0-\bar K^0),
  \qquad\qquad CP|K_1\rangle=|K_1\rangle~,
\end{equation}
\begin{equation}
K_2={1\over{\sqrt 2}}(K^0+\bar K^0),
  \qquad\qquad CP|K_2\rangle=-|K_2\rangle~,
\end{equation}
by 
a small admixture of the
other CP eigenstate:
\begin{equation}
K_{\rm S}={{K_1+\bar\varepsilon K_2}
\over{\sqrt{1+\mid\bar\varepsilon\mid^2}}},
\qquad
K_{\rm L}={{K_2+\bar\varepsilon K_1}
\over{\sqrt{1+\mid\bar\varepsilon\mid^2}}}\,.
\end{equation}

Since $\bar\varepsilon$ is $\ord(10^{-3})$, one has
 to a very good approximation:
\be\label{deltak1}
\Delta M_K = 2 \RE M_{12}, \qquad \Delta\Gamma_K=2 \RE \Gamma_{12}~,
\ee
where we have introduced the subscript K to stress that these formulae apply
only to the $K^0-\bar K^0$ system.

The 
$K_{\rm L}-K_{\rm S}$
mass difference is experimentally measured to be \cite{PDG}
\begin{equation}\label{DMEXP}
\Delta M_K=M(K_{\rm L})-M(K_{\rm S}) = 
(3.490\pm 0.006) \cdot 10^{-15} \gev\,.
\end{equation}
In the SM roughly $80\%$ of the measured $\Delta M_K$
is described by the real parts of the box diagrams with charm quark
and top quark exchanges, whereby the contribution of the charm exchanges
is by far dominant. 
The remaining $20 \%$ of the measured $\Delta M_K$ is attributed to long 
distance contributions which are difficult to estimate \cite{GERAR}.
Further information with the relevant references can be found in 
\cite{HNa}.
The situation with $\Delta \Gamma_K$ is rather different.
It is fully dominated by long distance effects. Experimentally
one has $\Delta\Gamma_K\approx-2 \Delta M_K$.

Generally to observe CP violation one needs an interference between
various amplitudes that carry complex phases. As these phases are
obviously convention dependent, the CP-violating effects depend only
on the differences of these phases. 
In particular
the parameter $\bar\varepsilon$  depends on the 
phase convention
chosen for $K^0$ and $\bar K^0$. Therefore it may not 
be taken as a physical measure of CP violation.
On the other hand $\RE~\bar\varepsilon$ and $r$, defined in
(\ref{bare3})  are independent of
phase conventions. In fact the departure of $r$ from 1
measures CP violation in the $K^0-\bar K^0$ mixing:
\be
r=1+\frac{2 |\Gamma_{12}|^2}{4 |M_{12}|^2+|\Gamma_{12}|^2}
    \IM\left(\frac{M_{12}}{\Gamma_{12}}\right)
\approx 1-\IM\left(\frac{\Gamma_{12}}{M_{12}}\right)~.
\ee

This type of CP violation can be best isolated in semi-leptonic
decays of the $ K_L$ meson. The non-vanishing
asymmetry $a_{\rm SL}(K_L)$:
\be\label{ASLK}
\frac{\Gamma(K_L\to \pi^-e^+\nu_e )-
                 \Gamma( K_L\to \pi^+e^-\bar\nu_e )}
{\Gamma(K_L\to \pi^-e^+\nu_e )+
                 \Gamma( K_L\to \pi^+e^-\bar\nu_e )}
          = \left(\IM\frac{\Gamma_{12}}{M_{12}}\right)_K
          = 2 \RE \bar\varepsilon
\ee
signals this type of CP violation.
Note that $a_{SL}(\rm K_L)$ is determined purely by the quantities
related to  $K^0-\bar K^0$ mixing. 
Specifically, it measures
the difference between the phases of $\Gamma_{12}$ and
$M_{12}$.

That a non--vanishing $a_{\rm SL}(K_L)$ is indeed a signal
of CP violation can also be understood in the following
manner. $K_L$, that should be a CP eigenstate $K_2$ in the case
of CP conservation, decays into CP conjugate final states with
different rates. As $\RE \bar\varepsilon>0$, $K_L$ prefers slightly
to decay into $\pi^-e^+\nu_e$ than $\pi^+e^-\bar\nu_e$.
This would not be possible in a CP-conserving world.

\subsection{The First Look at $\varepsilon$ and $\varepsilon'$}
Since a two pion final state is CP even while a three pion final state is CP
odd, $K_{\rm S}$ and $K_{\rm L}$ preferably decay to $2\pi$ and $3\pi$, 
respectively
via the following CP-conserving decay modes:
\begin{equation}
K_{\rm L}\to 3\pi {\rm ~~(via~K_2),}\qquad K_{\rm S}\to 2 
\pi {\rm ~~(via~K_1).}
\end{equation}
This difference is responsible for the large disparity in their
life-times. A factor of 579. 
However, $K_{\rm L}$ and $K_{\rm S}$ are not CP eigenstates and 
may decay with small branching fractions as follows:
\begin{equation}
K_{\rm L}\to 2\pi {\rm ~~(via~K_1),}\qquad K_{\rm S}\to 3 
\pi {\rm ~~(via~K_2).}
\end{equation}
This violation of CP is called {\it indirect} as it
proceeds not via explicit breaking of the CP symmetry in 
the decay itself but via the admixture of the CP state with opposite 
CP parity to the dominant one.
 The measure for this
indirect CP violation is defined as (I=isospin)
\begin{equation}\label{ek}
\varepsilon
\equiv {{A(K_{\rm L}\rightarrow(\pi\pi)_{I=0}})\over{A(K_{\rm 
S}\rightarrow(\pi\pi)_{I=0})}}.
\end{equation}

Following the derivation in \cite{CHAU83} one finds
\begin{equation}
\eps = \bar\varepsilon+i\xi= \frac{\exp(i \pi/4)}{\sqrt{2} \Delta M_K} \,
\left( \IM M_{12} + 2 \xi \RE M_{12} \right),
\quad\quad
\xi = \frac{\IM A_0}{\RE A_0}.
\label{eq:epsdef}
\end{equation}
The phase convention dependence of $\xi$ cancells
the one of $\bar\varepsilon$ so that $\varepsilon$
is free from this dependence. The isospin amplitude $A_0$ is defined
below.

The important point in the definition (\ref{ek}) is that only the
transition to $(\pi\pi)_{I=0}$ enters. The transition to
$(\pi\pi)_{I=2}$ is absent. This allows to remove a certain type
of CP violation that originates in decays only. Yet as 
$\varepsilon\not=\bar\varepsilon$ and only $\RE\varepsilon=
\RE\bar\varepsilon$, it is clear that $\varepsilon$ includes
a type of CP violation represented by $\IM\varepsilon$ which is
absent in the semileptonic asymmetry (\ref{ASLK}). We will
identify this type of CP violation in Section 2.7, where a more
systematic classification of different types of CP violation
will be given.

While {\it indirect} CP violation reflects the fact that the mass
eigenstates are not CP eigenstates, so-called {\it direct}
CP violation is realized via a 
direct transition of a CP odd to a CP even state: $K_2\to \pi\pi$.
A measure of such a direct CP violation in $K_L\to \pi\pi$ is characterized
by a complex parameter $\varepsilon'$  defined as
\be\label{eprime0}
\varepsilon'\equiv\frac{1}{\sqrt{2}}\left(\frac{A_{2,L}}{A_{0,S}}-
\frac{A_{2,S}}{A_{0,S}}\frac{A_{0,L}}{A_{0,S}}\right)
\ee
where $A_{I,L}\equiv A(K_L\to (\pi\pi)_I)$
and $A_{I,S}\equiv A(K_S\to (\pi\pi)_I)$.

This time the transitions to $(\pi\pi)_{I=0}$ and
$(\pi\pi)_{I=2}$ are included which allows to study CP violation in
the decay itself. We will discuss this issue in general terms
in Section 2.7. It is useful to cast (\ref{eprime0})
into 
\be\label{eprime}
\varepsilon'=\frac{1}{\sqrt{2}}\IM\left(\frac{A_2}{A_0}\right)
              \exp(i\Phi_{\varepsilon'}), 
   \qquad \Phi_{\varepsilon'}=\frac{\pi}{2}+\delta_2-\delta_0, 
\ee
where
the isospin amplitudes $A_I$ in $K\to\pi\pi$
decays are introduced through
\begin{equation}\label{ISO1} 
A(K^+\rightarrow\pi^+\pi^0)=\sqrt{3\over 2} A_2 e^{i\delta_2}~,
\end{equation}
\begin{equation}\label{ISO2}
A(K^0\rightarrow\pi^+\pi^-)=\sqrt{2\over 3} A_0 e^{i\delta_0}+ \sqrt{1\over
3} A_2 e^{i\delta_2}~,
\end{equation}
\begin{equation}\label{ISO3}
A(K^0\rightarrow\pi^0\pi^0)=\sqrt{2\over 3} A_0 e^{i\delta_0}-2\sqrt{1\over
3} A_2 e^{i\delta_2}\,.
\end{equation} 
Here the subscript $I=0,2$ denotes states with isospin $0,2$
equivalent to $\Delta I=1/2$ and $\Delta I = 3/2$ transitions,
respectively, and $\delta_{0,2}$ are the corresponding strong phases. 
The weak CKM phases are contained in $A_0$ and $A_2$.
The isospin amplitudes $A_I$ are complex quantities which depend on
phase conventions. On the other hand, $\varepsilon'$ measures the 
difference between the phases of $A_2$ and $A_0$ and is a physical
quantity.
The strong phases $\delta_{0,2}$ can be extracted from $\pi\pi$ scattering. 
Then $\Phi_{\varepsilon'}\approx \pi/4$. See \cite{Meissner} for more details.

Experimentally $\varepsilon$ and $\varepsilon'$
can be found by measuring the ratios
\begin{equation}
\eta_{00}={{A(K_{\rm L}\to\pi^0\pi^0)}\over{A(K_{\rm S}\to\pi^0\pi^0)}},
            \qquad
  \eta_{+-}={{A(K_{\rm L}\to\pi^+\pi^-)}\over{A(K_{\rm S}\to\pi^+\pi^-)}}.
\end{equation}

Indeed, assuming $\varepsilon$ and $\varepsilon'$ to be small numbers one
finds
\be
\eta_{00}=\varepsilon-{{2\varepsilon'}\over{1-\sqrt{2}\omega}}
            ,~~~~
  \eta_{+-}=\varepsilon+{{\varepsilon'}\over{1+\omega/\sqrt{2}}}
\end{equation}
where $\omega=\RE A_2/\RE A_0=0.045$.
In the absence of direct CP violation $\eta_{00}=\eta_{+-}$.
The ratio ${\varepsilon'}/{\varepsilon}$  can then be measured through
\begin{equation}\label{BASE}
\RE(\epe)=\frac{1}{6(1+\omega/\sqrt{2})}
\left(1-\left|{{\eta_{00}}\over{\eta_{+-}}}\right|^2\right)~.
\end{equation}

\subsection{Basic Formula for $\eps$}
            \label{subsec:epsformula}
With all this information at hand one can derive a formula for $\varepsilon$
which can be efficiently used in pheneomenological applications.
As this derivation has been presented in detail in \cite{Erice}, we will be 
very brief here.

Calculating the box diagrams of fig. \ref{L:9} and including
leading and next-to-leading QCD corrections one finds
\begin{equation}
M_{12} = D_{\varepsilon}
\left[ {\lambda_c^*}^2 \eta_1 S_0(x_c) + {\lambda_t^*}^2 \eta_2 S_0(x_t) +
2 {\lambda_c^*} {\lambda_t^*} \eta_3 S_0(x_c, x_t) \right],
\label{eq:M12K}
\end{equation}
\be
D_{\varepsilon}=\frac{G_{\rm F}^2}{12 \pi^2} F_K^2 \hat B_K m_K \mw^2
\ee
where $F_K=160~\mev$ is the $K$-meson decay constant and $m_K$
the $K$-meson mass. 
Next, the renormalization group 
invariant parameter $\hat B_K$ is defined by
\begin{equation}
\hat B_K = B_K(\mu) \left[ \alpha_s^{(3)}(\mu) \right]^{-2/9} \,
\left[ 1 + \frac{\alpha_s^{(3)}(\mu)}{4\pi} J_3 \right]~,
\label{eq:BKrenorm}
\end{equation}
\begin{equation}
\langle \bar K^0| (\bar s d)_{V-A} (\bar s d)_{V-A} |K^0\rangle
\equiv \frac{8}{3} B_K(\mu) F_K^2 m_K^2
\label{eq:KbarK}
\end{equation}
where $\alpha_s^{(3)}$ is the strong coupling constant
in an effective three flavour theory and $J_3=1.895$ in the NDR scheme 
\cite{BJW90}. The CKM factors are given by $\lambda_i = V_{is}^* V_{id}^{}$ 
and the functions $S_0$  by ($x_i=m^2_i/\mw^2$)
\begin{equation}\label{S0}
 S_0(x_t)=2.39~\left(\frac{\mt}{167\gev}\right)^{1.52},
\quad\quad S_0(x_c)=x_c,
\ee
\begin{equation}\label{BFF}
S_0(x_c, x_t)=x_c\left[\ln\frac{x_t}{x_c}-\frac{3x_t}{4(1-x_t)}-
 \frac{3 x^2_t\ln x_t}{4(1-x_t)^2}\right].
\end{equation}

Short-distance NLO QCD effects are described through the correction
factors $\eta_1$, $\eta_2$, $\eta_3$ \cite{HNa,BJW90,HNb,Nierste}:
\begin{equation}
\eta_1=(1.32\pm 0.32) \left[\frac{1.30\gev}{m_c(m_c)}\right]^{1.1},\quad
\eta_2=0.57\pm 0.01,\quad
  \eta_3=0.47\pm0.05~.
\end{equation}

To proceed further we neglect the last term in (\eqn{eq:epsdef}) as it
 constitutes at most a 2\,\% correction to $\eps$. This is justified
in view of other uncertainties, in particular those connected with
$\hat B_K$.
Inserting (\eqn{eq:M12K}) into (\eqn{eq:epsdef}) we find
\begin{equation}
\eps=C_{\eps} \hat B_K \IM\lambda_t \left\{
\RE\lambda_c \left[ \eta_1 S_0(x_c) - \eta_3 S_0(x_c, x_t) \right] -
\RE\lambda_t \eta_2 S_0(x_t) \right\} e^{i \pi/4}\,,
\label{eq:epsformula}
\end{equation}
where the numerical constant $C_\eps$ is given by
\begin{equation}
C_\eps = \frac{G_{\rm F}^2 F_K^2 m_K \mw^2}{6 \sqrt{2} \pi^2 \Delta M_K}
       = 3.837 \cdot 10^4 \, .
\label{eq:Ceps}
\end{equation}
Comparing (\eqn{eq:epsformula}) with the
experimental value for $\eps$ \cite{PDG}
\begin{equation}\label{eexp}
\varepsilon_{exp}
=(2.280\pm0.013)\cdot10^{-3}\;\exp{i\Phi_{\varepsilon}},
\qquad \Phi_{\varepsilon}={\pi\over 4},
\end{equation}
one obtains a constraint on the unitarity triangle in 
fig.~\ref{fig:utriangle}. 
See Section 3.

\subsection{Express Review of $B_{d,s}^0$-$\bar B_{d,s}^0$ Mixing}
The flavour eigenstates in this case are
\be\label{fl}
B^0_d=(\bar bd),\qquad
\bar B^0_d=(b \bar d),\qquad
B^0_s=(\bar bs),\qquad
\bar B^0_s=( b \bar s)~.
\ee
They mix via the box diagrams in fig.~\ref{L:9} with $s$ replaced
by $b$ in the case of $B_{d}^0$-$\bar B_{d}^0$ mixing.
In the case of $B_{s}^0$-$\bar B_{s}^0$ mixing also $d$ has to be replaced
by $s$.

Dropping the subscripts $(d,s)$ for a moment, it is customary to
denote the mass eigenstates by
\be\label{HL}
B_H=p B^0+q \bar B^0, \qquad B_L=p B^0-q \bar B^0,
\ee
\be\label{pq}
p=\frac{1+\bar\varepsilon_B}{\sqrt{2(1+|\bar\varepsilon_B|^2)}},
\qquad
q=\frac{1-\bar\varepsilon_B}{\sqrt{2(1+|\bar\varepsilon_B|^2)}},
\ee
with $\bar\varepsilon_B$ corresponding to $\bar\varepsilon$ in
the $K^0-\bar K^0$ system. Here ``H'' and ``L'' denote 
{\it Heavy} and {\it Light} respectively. As in the $B^0-\bar B^0$ 
system one has $\Delta\Gamma\ll\Delta M$, 
 it is more suitable to distinguish the mass eigenstates by their 
masses than the corresponding life-times.

The strength of the $B^0_{d,s}-\bar B^0_{d,s}$ mixings
is described by the mass differences
\begin{equation}
\Delta M_{d,s}= M_H^{d,s}-M_L^{d,s}~.
\end{equation}
In contrast to $\Delta M_K$ , in this case the long distance contributions
are estimated to be very small and $\Delta M_{d,s}$ is very well
approximated by the relevant box diagrams. 
Moreover, due $m_{u,c}\ll m_t$ 
only the top sector is relevant.

 $\Delta M_{d,s}$ can be expressed
in terms of the off-diagonal element in the neutral B-meson mass matrix
by using the formulae developed previously for the K-meson system.
One finds
\begin{equation}
\Delta M_q= 2 |M_{12}^{(q)}|, \qquad
\Delta \Gamma_q=2 \frac{\RE(M_{12}\Gamma_{12}^*)}{|M_{12}|} \ll\Delta M_q, \qquad
 q=d,s.
\label{eq:xdsdef}
\end{equation}
These formulae differ from (\ref{deltak1}) because in the
B-system $\Gamma_{12}\ll M_{12}$.

We also have
\be\label{q/p}
\frac{q}{p}=
\frac{2 M^*_{12}-i\Gamma^*_{12}}{\Delta M-i\frac{1}{2}\Delta\Gamma}
=\frac{M_{12}^*}{|M_{12}|}
\left[1-\frac{1}{2}\IM\left(\frac{\Gamma_{12}}{M_{12}}\right)\right]
\ee
where higher order terms in the small quantity $\Gamma_{12}/M_{12}$
have been neglected.

As $\IM(\Gamma_{12}/M_{12})< \ord(10^{-3})$,
\bi
\item
The semileptonic asymmetry $a_{\rm SL}(B)$ discussed a few
pages below is even smaller than $a_{\rm SL}(K_L)$. Typically 
$\ord(10^{-4})$. These are bad news.
\item
The ratio $q/p$ is a pure phase to an excellent approximation.
These are very good news as we will see below.
\ei
Inspecting the relevant box diagrams we find
\be
(M_{12}^*)_d \propto (V_{td}V_{tb}^*)^2~,
\qquad
(M_{12}^*)_s \propto (V_{ts}V_{tb}^*)^2~.
\ee
Now, from Section 1 we know that
\be
V_{td}=\vtd e^{-i\beta}, \qquad
V_{ts}=-\vts e^{-i\beta_s}
\ee
with $\beta_s=\ord(10^{-2})$. Consequently to an excellent approximation 
\be\label{pureph}
\left(\frac{q}{p}\right)_{d,s}= e^{i2\phi_M^{d,s}},
\qquad
\phi^d_M=-\beta, \qquad \phi^s_M=-\beta_s,
\ee
with $\phi_M^{d,s}$ given entirely by the weak phases in the
CKM matrix.

\subsection{Basic Formulae for $\Delta M_{d,s}$}
            \label{subsec:BBformula}
The formulae for $\Delta M_{d,s}$ have been derived in \cite{Erice} with 
the result
\begin{equation}
\Delta M_q = \frac{G_{\rm F}^2}{6 \pi^2} \eta_B m_{B_q} 
(\hat B_{B_q} F_{B_q}^2 ) \mw^2 S_0(x_t) |V_{tq}|^2,
\label{eq:xds}
\end{equation}
where $F_{B_q}$ is the $B_q$-meson decay constant, $\hat B_q$
renormalization group invariant parameters defined
in analogy to (\ref{eq:BKrenorm}) and (\ref{eq:KbarK}) and $\eta_B$ stands 
for short distance QCD corrections \cite{BJW90,UKJS}
\begin{equation}
\eta_B=0.55\pm0.01.
\end{equation}
Using (\ref{eq:xds}) we obtain two useful formulae
\begin{equation}\label{DMD}
\Delta M_d=
0.50/{\rm ps}\cdot\left[ 
\frac{\sqrt{\hat B_{B_d}}F_{B_d}}{230\mev}\right]^2
\left[\frac{\mtb(\mt)}{167\gev}\right]^{1.52} 
\left[\frac{\vtd}{7.8\cdot10^{-3}} \right]^2 
\left[\frac{\eta_B}{0.55}\right]  
\end{equation}
and
\begin{equation}\label{DMS}
\Delta M_{s}=
17.2/{\rm ps}\cdot\left[ 
\frac{\sqrt{\hat B_{B_s}}F_{B_s}}{260\mev}\right]^2
\left[\frac{\mtb(\mt)}{167\gev}\right]^{1.52} 
\left[\frac{\vts}{0.040} \right]^2
\left[\frac{\eta_B}{0.55}\right] \,.
\end{equation}

\subsection{Classification of CP Violation}
\subsubsection{Preliminaries}
We have mentioned in Section 1 that due to the presence of hadronic
matrix elements, various decay amplitudes contain large theoretical
uncertainties. It is of interest to investigate which measurements
of CP-violating effects do not suffer from hadronic uncertainties.
To this end it is useful to make a classification of CP-violating
effects that is more transparent than the division into the
{\it indirect} and {\it direct} CP violation considered so far.
A nice detailed presentation has been given by Nir \cite{REV}.

Generally complex phases may enter particle--antiparticle mixing
and the decay process itself. It is then natural to consider
three types of CP violation:
\bi
\item
CP Violation in Mixing
\item
CP Violation in Decay
\item
CP Violation in the Interference of Mixing and Decay
\ei

As the phases in mixing and decay are convention dependent,
the CP-violating effects depend only
on the differences of these phases. This is clearly seen in
the classification given below.

\subsubsection{CP Violation in Mixing}
This type of CP violation can be best isolated in semi-leptonic
decays of neutral B and K mesons. We have discussed the asymmetry
$a_{SL}(K_L)$ before. In the case of B decays the non-vanishing
asymmetry $a_{SL}(B)$ (we suppress the indices $(d,s)$),
\be\label{ASLB}
\frac{\Gamma(\bar B^0(t)\to l^+\nu X)-
                 \Gamma( B^0(t)\to l^-\bar\nu X)}
                {\Gamma(\bar B^0(t)\to l^+\nu X)+
                 \Gamma( B^0(t)\to l^-\bar\nu X)}
            =\frac{1-|q/p|^4}{1+|q/p|^4}
          = \left(\IM\frac{\Gamma_{12}}{M_{12}}\right)_B
\ee
signals this type of CP violation. Here $\bar B^0(0)=\bar B^0$, 
$B^0(0)= B^0$. For $t\not=0$ the formulae analogous to 
(\ref{SCH}) should be used.
Note that the final states in (\ref{ASLB}) contain ``wrong charge''
leptons and can only be reached in the presence of $B^0-\bar B^0$
mixing. That is one studies effectively the difference between the
rates for $\bar B^0\to B^0\to l^+\nu X$
and $ B^0 \to \bar B^0 \to l^-\bar\nu X$. 
As the phases in
the transitions $B^0 \to \bar B^0$ and $\bar B^0 \to B^0$ 
differ from each other, a non-vanishing CP asymmetry follows.
Specifically $a_{\rm SL}(B)$ measures
the difference between the phases of $\Gamma_{12}$ and
$M_{12}$.

As $M_{12}$ and 
in particular $\Gamma_{12}$ suffer from large hadronic uncertainties,
no precise extraction of CP-violating phases from this type of CP
violation can be expected.  
Moreover as $q/p$ is almost a pure phase, see (\ref{q/p}) and 
(\ref{pureph}), the
asymmetry is very small and very difficult to measure.

\subsubsection{CP Violation in Decay}
This type of CP violation is best isolated in charged B and charged K
decays as mixing effects do not enter here. However, it can also
be measured in neutral B and K decays. The relevant asymmetry is
given by
\be\label{ADECAY}
a^{\rm decay}_{f^\pm}=\frac{\Gamma(B^+\to f^+)-\Gamma(B^-\to f^-)}
                          {\Gamma(B^+\to f^+)+\Gamma(B^-\to f^-)}
=\frac{1-|\bar A_{f^-}/A_{f^+}|^2}{1+| \bar A_{f^-}/A_{f^+}|^2}
\ee
where
\be\label{AH}
A_{f^+}=\langle f^+|{\cal H}^{\rm weak}| B^+\rangle,
\qquad
\bar A_{f^-}=\langle f^-|{\cal H}^{\rm weak}| B^-\rangle~.
\ee
For this asymmetry to be non-zero one needs at least two different 
contributions with different {\it weak} ($\phi_i$) and {\it strong}
($\delta_i$) phases. These could be for instance two tree diagrams,
two penguin diagrams or one tree and one penguin. Indeed writing the
decay amplitude $A_{f^+}$ and its CP conjugate $\bar A_{f^-}$ as
\be\label{AMPL}
A_{f^+}=\sum_{i=1,2} A_i e^{i(\delta_i+\phi_i)},
\qquad
\bar A_{f^-}=\sum_{i=1,2} A_i e^{i(\delta_i-\phi_i)},
\ee
with $A_i$ being real, one finds 
\be\label{BDECAY}
a^{\rm decay}_{f^\pm}=\frac{ -2 A_1 A_2 \sin(\delta_1-\delta_2)
\sin(\phi_1-\phi_2)}{A_1^2+A_2^2+2 A_1 A_2 \cos(\delta_1-\delta_2)
\cos(\phi_1-\phi_2)}~.
\ee
The sign of strong phases $\delta_i$ is the same for $A_{f^+}$
and $\bar A_{f^-}$ because CP is conserved by strong interactions.
The weak phases have opposite signs. 

The presence of hadronic uncertainties in $A_i$ and 
of strong phases $\delta_i$ complicates the extraction of the
phases $\phi_i$ from data. An example of this type of
CP violation in K decays is $\varepsilon'$. We will demonstrate
this below.
\subsubsection{CP Violation in the Interference of Mixing and Decay}
This type of CP violation is only possible in neutral B and K
decays. We will use B decays for illustration suppressing the
subscripts $d$ and $s$. Moreover, we set $\Delta\Gamma=0$. Formulae
with $\Delta\Gamma\not =0$ can be found in \cite{BF97,REV}.

 Most interesting are the decays into final states which
are CP-eigenstates. Then a time dependent asymmetry defined by
\be\label{TASY}
a_{CP}(t,f)=\frac{\Gamma(B^0(t)\to f)-
                 \Gamma( \bar B^0(t)\to f)}
                {\Gamma(B^0(t)\to f)+
                 \Gamma( \bar B^0(t)\to f)}
\ee
is given by
\begin{equation}\label{e8}
a_{CP}(t,f)=
a^{\rm decay}_{CP}(f)\cos(\Delta M
t)+a^{\rm int}_{CP}(f)\sin(\Delta M t)
\end{equation}
where we have separated the {\it decay} CP-violating contributions 
from those describing CP violation in the interference of
mixing and decay:
\begin{equation}\label{e9}
a^{\rm decay}_{CP}(f)=\frac{1-\left\vert\xi_f\right\vert^2}
{1+\left\vert\xi_f\right\vert^2}\equiv C_f,
\quad
a^{\rm int}_{CP}(f)=\frac{2\mbox{Im}\xi_f}{1+
\left\vert\xi_f\right\vert^2}\equiv -S_f~.
\end{equation}
Here $C_f$ and $S_f$ are popular notations found in the recent literature.
The later type of CP violation is sometimes called the
{\it mixing-induced} CP violation. 
The quantity $\xi_f$ containing  all the information
needed to evaluate the asymmetries (\ref{e9}) 
is given by
\begin{equation}\label{e11}
\xi_f=\frac{q}{p}\frac{A(\bar B^0\to f)}{A(B^0 \to f)}=
\exp(i2\phi_M)\frac{A(\bar B^0\to f)}{A(B^0 \to f)}
\end{equation}
with $\phi_M$, introduced in (\ref{pureph}), 
denoting the weak phase in the $B^0-\bar B^0$ mixing.
 $A(B^0 \to f)$ and $A(\bar B^0 \to f)$ are  decay amplitudes. 
The time dependence of $a_{CP}(t,f)$ allows to extract
$a^{\rm decay}_{CP}$ and $a^{\rm int}_{CP}$ as
coefficients of $\cos(\Delta M t)$ and $\sin(\Delta M t)$,
respectively.

Generally several decay mechanisms with different weak and
strong phases can contribute to $A(B^0 \to f)$. These are
tree diagram (current-current) contributions, QCD penguin
contributions and electroweak penguin contributions. If they
contribute with similar strength to a given decay amplitude
the resulting CP asymmetries suffer from hadronic uncertainies
related to matrix elements of the relevant operators $Q_i$.
The situation is then analogous to the class just discussed.
Indeed
\be\label{ratiocp}
\frac{A(\bar B^0\to f)}{A(B^0 \to f)}=-\eta_f
\left[\frac{A_T e^{i(\delta_T-\phi_T)}+A_P e^{i(\delta_P-\phi_P)}}
{A_T e^{i(\delta_T+\phi_T)}+A_P e^{i(\delta_P+\phi_P)}}\right]
\ee
with $\eta_f=\pm 1$ being the CP-parity of the final state,
depends on strong phases $\delta_{T,P}$ and hadronic matrix
elements present in $A_{T,P}$. Thus the measurement of the
asymmetry does not allow a clean determination of the weak
phases $\phi_{T,P}$. The minus sign in (\ref{ratiocp}) follows
from our CP phase convention $ CP |B^0\rangle= -|\bar B^0\rangle$,
 that has also been used in writing the phase factor in (\ref{e11}).
Only $\xi$ is phase convention independent. See Section 8.4.1 of 
\cite{BF97} for details.

An interesting case arises when a single mechanism dominates the 
decay amplitude or the contributing mechanisms have the same weak 
phases. Then the hadronic matrix elements and strong phases drop out and
\be\label{cp}
\frac{A(\bar B^0\to f)}{A(B^0 \to f)}=-\eta_f e^{-i2\phi_D}
\ee
is a pure phase with $\phi_D$ being the weak phase in 
$A(B^0 \to f)$.
Consequently
\begin{equation}\label{e111}
\xi_f=-\eta_f\exp(i2\phi_M) \exp(-i 2 \phi_D),
\qquad
\mid \xi_f \mid^2=1~.
\end{equation}
In this particular case 
$a^{\rm decay}_{CP}(f)=C_f$
vanishes and the CP asymmetry is given entirely
in terms of the weak phases $\phi_M$ and $\phi_D$:
\begin{equation}\label{simple}
a_{CP}(t,f)= \IM\xi_f \sin(\Delta Mt) \qquad
\IM\xi_f=\eta_f \sin(2\phi_D-2\phi_M)=-S_f~.
\end{equation}
Thus the corresponding measurement of weak phases is free from
hadronic uncertainties. A well known example is the decay
$B_d\to \psi K_S$. Here $\phi_M=-\beta$ and $\phi_D=0$. As
in this case $\eta_f=-1$,  we find
\begin{equation}
a_{CP}(t,f)= -\sin(2\beta) \sin(\Delta Mt), \qquad S_f=\sin(2\beta)
\end{equation}
which allows a very clean measurement of the angle $\beta$ in the
unitarity triangle. We will discuss other examples in Section 5.

We observe that the asymmetry $a_{CP}(t,f)$ measures directly the
difference between the phases of $B^0-\bar B^0$-mixing $(2\phi_M)$
and of the decay amplitude $(2\phi_D)$. This tells us immediately 
that we are dealing with the interference of mixing and decay.
As $\phi_M$ and $\phi_D$
are phase convention dependent quantities, only their
difference is physical, it is
impossible to state on the basis of a single asymmetry 
whether CP violation takes place in the decay or in the mixing.
To this end at least two asymmetries for $B^0 (\bar B^0)$
decays to different final states $f_i$ have to be measured.
As $\phi_M$ does not depend on the final state, 
$\IM\xi_{f_1}\not=\IM\xi_{f_2}$ is a signal of CP violation
in the decay. 

We will see in Section 5 that the ideal situation presented above 
does not always take place and two or more different mechanism with 
different weak and strong phases contribute to the CP asymmetry. One
finds then
\begin{equation}\label{e8a}
a_{CP}(t,f)=C_f\cos(\Delta Mt)-S_f\sin(\Delta M t),
\end{equation}
\be\label{Cf}
C_f=-2 r \sin(\phi_1-\phi_2)\sin(\delta_1-\delta_2)~,
\ee
\be\label{Sf}
S_f=-\eta_f\left[\sin 2(\phi_1-\phi_M)+
2r \cos 2(\phi_1-\phi_M) \sin(\phi_1-\phi_2)\cos(\delta_1-\delta_2)\right]
\ee
where 
$r=A_2/A_1\ll 1$
and $\phi_i$ and $\delta_i$ are weak and strong phases, respectively.
For $r=0$ the previous formulae are obtained.

In the case of K decays, this type of CP violation can be
cleanly measured in the rare decay $K_L\to\pi^0\nu\bar\nu$.
Here the difference between the weak phase in the $K^0-\bar K^0$
mixing and in the decay $\bar s \to \bar d \nu\bar\nu$ matters.

We can now compare the two classifications of different types
of CP violation. CP violation in mixing is a manifestation
of indirect CP violation. CP violation in decay is a manifestation
of direct CP violation. CP violation in interference of mixing
and decay contains elements of both the indirect and direct CP
violation.

It is clear from this discussion that only in the case of the
third type of CP violation there are possibilities to measure directly weak
phases without hadronic uncertainties and moreover without invoking sophisticated 
methods. This takes place provided
a single mechanism (diagram) is responsible for the decay or the
contributing decay mechanisms have the same weak phases.
However, we will see in Section 5 that there are other strategies, involving 
also decays to CP non-eigenstates, that provide clean measurements of the 
weak phases.

\subsubsection{Another Look at $\varepsilon$ and $\varepsilon'$}
Let us finally investigate what type of CP violation is
represented by $\varepsilon$ and $\varepsilon'$.
Here instead of different mechanisms it is sufficient to talk
about different isospin amplitudes.

In the case of $\varepsilon$, CP violation in decay is not
possible as only the isospin amplitude $A_0$ is involved.
See (\ref{ek}). We know also that only $\RE\varepsilon=
\RE\bar\varepsilon$ is related to CP violation in mixing.
Consequently:
\bi
\item 
$\RE\varepsilon$ represents CP violation in mixing,
\item
$\IM\varepsilon$ represents CP violation in the
interference of mixing and decay.
\ei

In order to analyze the case of $\varepsilon'$ we use the formula
(\ref{eprime}) to find
\be\label{ree}
\RE\,\varepsilon'=-\frac{1}{\sqrt{2}}\left\vert\frac{A_2}{A_0}\right\vert
\sin(\phi_2-\phi_0)\sin(\delta_2-\delta_0)
\ee
\be\label{iee}
\IM\,\varepsilon'=\frac{1}{\sqrt{2}}\left\vert\frac{A_2}{A_0}\right\vert
\sin(\phi_2-\phi_0)\cos(\delta_2-\delta_0)~.
\ee  
Consequently:
\bi
\item 
$\RE~\varepsilon'$ represents CP violation in decay as it is only
non zero provided simultaneously $\phi_2\not=\phi_0$ and 
$\delta_2\not=\delta_0$.
\item
$\IM~\varepsilon'$ exists even for $\delta_2=\delta_0$ but as
it requires $\phi_2\not=\phi_0$ it represents CP violation in 
decay as well.
\ei
Experimentally  $\delta_2\not=\delta_0$.
Within the SM, $\phi_2$ and $\phi_0$ are connected with
electroweak penguins and QCD penguins, respectively.
We will see in Section 4 that these  phases differ from
each other so that a nonvanishing $\varepsilon'$ is obtained.

\section{Standard Analysis of the Unitarity Triangle (UT)}\label{UT-Det}
\subsection{General Procedure}
After these general discussion of basic concepts let us concentrate on
the standard analysis of the Unitarity Triangle (see 
fig. \ref{fig:utriangle}) within the SM. 
A very detailed description of this analysis with the participation of 
the leading experimentalists and theorists in this field 
can be found in  \cite{CERNCKM}.
 
Setting $\lambda=\vus=0.224$, the analysis
proceeds in the following five steps:

{\bf Step 1:}

{}From  $b\to c$ transition in inclusive and exclusive 
leading B-meson decays
one finds $\vcb$ and consequently the scale of the UT:
\begin{equation}
\vcb\quad \Longrightarrow\quad\lambda \vcb= \lambda^3 A~.
\end{equation}

{\bf Step 2:}

{}From  $b\to u$ transition in inclusive and exclusive $B$ meson decays
one finds $\vub$ and consequently using (\ref{2.94}) 
the side $CA=R_b$ of the UT:
\begin{equation}\label{rb}
\left| \frac{V_{ub}}{V_{cb}} \right|
 \quad\Longrightarrow \quad R_b=\sqrt{\bar\varrho^2+\bar\eta^2}=
4.35 \cdot \left| \frac{V_{ub}}{V_{cb}} \right|~.
\end{equation}

{\bf Step 3:}

{}From the experimental value of $\varepsilon_K$ in (\ref{eexp})   
and the formula (\ref{eq:epsformula}) rewritten in terms of Wolfenstein 
parameters
one derives  
the constraint on $(\bar \varrho, \bar\eta)$ \cite{WARN}
\begin{equation}\label{100}
\bar\eta \left[(1-\bar\varrho) A^2 \eta_2 S_0(x_t)
+ P_c(\varepsilon) \right] A^2 \hat B_K = 0.187,
\end{equation}
where
\begin{equation}\label{102}
P_c(\varepsilon) = 
\left[ \eta_3 S_0(x_c,x_t) - \eta_1 x_c \right] \frac{1}{\lambda^4},
\qquad
x_i=\frac{m^2_i}{\mw^2}
\end{equation}
with all symbols defined in the previous Section and
$P_c(\varepsilon)=0.29\pm0.07$ \cite{Nierste} summarizing the contributions
of box diagrams with two charm quark exchanges and the mixed 
charm-top exchanges.  

As seen in fig.~\ref{L:10}, equation (\ref{100}) specifies 
a hyperbola in the $(\bar \varrho, \bar\eta)$
plane.
The position of the hyperbola depends on $\mt$, $|V_{cb}|=A \lambda^2$
and $\hat B_K$. With decreasing $\mt$, $|V_{cb}|$ and $\hat B_K$ it
moves away from the origin of the
$(\bar\varrho,\bar\eta)$ plane. 

\begin{figure}[hbt]
  \vspace{0.10in} \centerline{
\begin{turn}{-90}
  \mbox{\epsfig{file=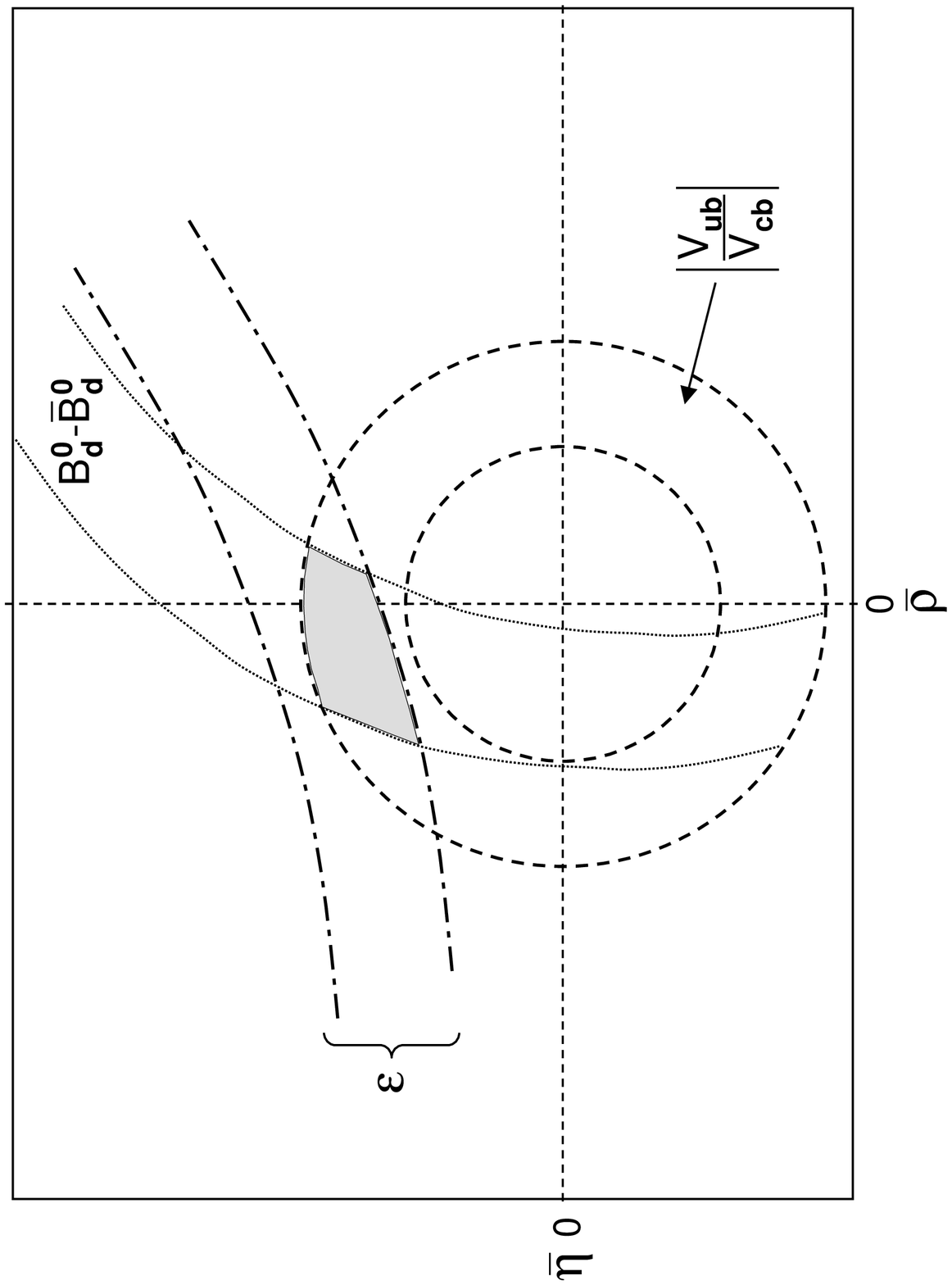,width=0.5\linewidth}}
\end{turn}
} \vspace{-0.18in}
\caption[]{Schematic determination of the Unitarity Triangle.
\label{L:10}}
 \end{figure}
{\bf Step 4:}

{}From the measured $\Delta M_d$ 
and the formula (\ref{DMD}),
the side $AB=R_t$ of the UT can be determined:
\begin{equation}\label{106}
 R_t= \frac{1}{\lambda}\frac{|V_{td}|}{\vcb} = 0.85 \cdot
\left[\frac{|V_{td}|}{7.8\cdot 10^{-3}} \right] 
\left[ \frac{0.041}{\vcb} \right],
\end{equation}
\begin{equation}\label{VT}
\vtd=
7.8\cdot 10^{-3}\left[ 
\frac{230\mev}{\sqrt{\hat B_{B_d}}F_{B_d}}\right]
\left[\frac{167~GeV}{\mtb(\mt)} \right]^{0.76} 
\left[\frac{\Delta M_d}{0.50/{\rm ps}} \right ]^{0.5} 
\sqrt{\frac{0.55}{\eta_B}}
\end{equation}
with all symbols defined in the previous Section.
$\mtb(\mt)=(167\pm 5)$ GeV.
Note that $R_t$ suffers from additional uncertainty in $\vcb$,
which is absent in the determination of $\vtd$ this way. 
The constraint in the $(\bar\varrho,\bar\eta)$ plane coming from
this step is illustrated in fig.~\ref{L:10}.

{\bf Step 5:}

{}The measurement of  $\Delta M_s$
together with $\Delta M_d$  allows to determine $R_t$ in a different
manner:
\be\label{Rt}
R_t=0.90~\left[\frac{\xi}{1.24}\right] \sqrt{\frac{18.4/ps}{\Delta M_s}} 
\sqrt{\frac{\Delta M_d}{0.50/ps}},
\qquad
\xi = 
\frac{\sqrt{\hat B_{B_s}}F_{B_s} }{ \sqrt{\hat B_{B_d}}F_{B_d}}.
\ee
One should 
note that $\mt$ and $|V_{cb}|$ dependences have been eliminated this way
 and that $\xi$ should in principle 
contain much smaller theoretical
uncertainties than the hadronic matrix elements in $\Delta M_d$ and 
$\Delta M_s$ separately.  

The main uncertainties in these steps originate in the theoretical 
uncertainties in  $\hat B_K$ and 
$\sqrt{\hat B_d}F_{B_d}$ and to a lesser extent in $\xi$ 
\cite{CERNCKM}: 
\be
\hat B_K=0.86\pm0.15, \quad  \sqrt{\hat B_d}F_{B_d}=(235^{+33}_{-41})~MeV,
\quad \xi=1.24\pm 0.08~.
\ee
Also the uncertainties due to $\vub$ in step 2  
are substantial. The QCD sum rules results for  
the parameters in question are similar and can be found in 
\cite{CERNCKM}. 
Finally \cite{CERNCKM}
\be
\Delta M_d=(0.503\pm0.006)/{\rm ps}, \qquad 
\Delta M_s>14.4/{\rm ps}~~ {\rm at }~~ 95\%~{\rm C.L.}
\ee 
\subsection{The Angle \boldmath{$\beta$} from \boldmath{$B_d\to \psi K_S$}}
One of the highlights of the year 2002 were the considerably improved 
measurements of 
$\sin2\beta$ by means of the time-dependent CP asymmetry
\be\label{asy}
a_{\psi K_S}(t)\equiv -a_{\psi K_S}\sin(\Delta M_d t)=
-\sin 2 \beta \sin(\Delta M_d t)~.
\ee
The BaBar \cite{BaBar} and Belle \cite{Belle} collaborations find
\begin{displaymath}\label{sinb}
(\sin 2\beta)_{\psi K_S}=\left\{
\begin{array}{ll}
0.741\pm 0.067 \, \mbox{(stat)} \pm0.033 \, \mbox{(syst)} & \mbox{(BaBar)}\\
0.719\pm 0.074 \, \mbox{(stat)} \pm0.035  \, \mbox{(syst)} & \mbox{(Belle).}
\end{array}
\right.
\end{displaymath}
Combining these results with earlier measurements by CDF 
$(0.79^{+0.41}_{-0.44})$, ALEPH $(0.84^{+0.82}_{-1.04}\pm 0.16)$ and OPAL 
gives the grand average \cite{NIR02}
\be
(\sin 2\beta)_{\psi K_S}=0.734\pm 0.054~.
\label{ga}
\ee
This is a mile stone in the field of CP violation and in the tests of the
SM as we will see in a moment. Not only violation of this symmetry has been 
confidently established 
in the B system, but also its size has been measured very accurately.
Moreover in contrast to the five constraints listed above, the determination 
of the angle $\beta$ in this manner is theoretically very clean.
\subsection{Unitarity Triangle 2003}
We are now in the position to combine all these constraints in order to 
construct the unitarity triangle and determine various quantities of interest.
In this context the important issue is the error analysis of these formulae, 
in particular the treatment of theoretical uncertainties. In the 
literature the most popular are the 
Bayesian approach \cite{C00} and the frequentist approach \cite{FREQ}. 
For the PDG analysis see \cite{PDG}.
A critical comparison of these and other methods can be found in 
\cite{CERNCKM}.
I can recommend this reading. 

In fig.~\ref{fig:figmfv} we show the result of the recent update of 
an analysis in collaboration 
with Parodi and Stocchi \cite{BUPAST} that uses 
the Bayesian approach. The results presented below are very close to 
the ones presented in \cite{CERNCKM} that was led by my Italian collaborators. 
The allowed region for $(\bar\varrho,\bar\eta)$ 
is the area inside the smaller ellipse.
We observe that the region
$\bar\varrho<0$ is disfavoured by the lower bound on
$\Delta M_s$.
It is clear
from this figure that the measurement of $\Delta M_s$
giving $R_t$ through (\ref{Rt}) will have a large impact
on the plot in fig.~\ref{fig:figmfv}.  

\begin{figure}[htb!]
\begin{center}
{\epsfig{figure=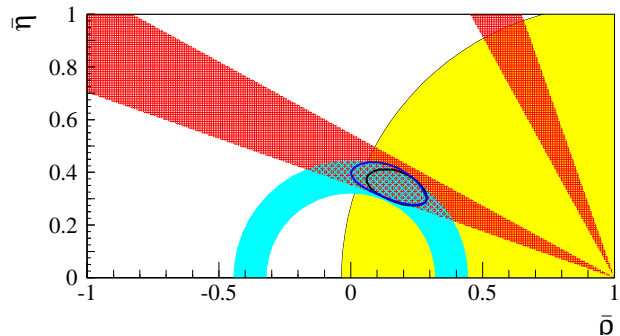,height=5cm}}
\caption[]{The allowed 95$\%$ regions in the 
$(\bar\varrho,\bar\eta)$ plane in the SM (narrower region) and in the 
MFV models (broader region) from the update of \cite{BUPAST}.
The individual 95$\%$ regions for the constraint from 
$\sin 2 \beta$, $\Delta M_s$ and $R_b$ are also shown.  
}
\label{fig:figmfv}
\end{center}
\end{figure}

The ranges for various quantities that result from this analysis 
are given in the SM column of table~\ref{mfv}. The UUT column 
will be discussed in Section 7. The SM results follow from 
the five steps listed above and (\ref{ga}) implying an impressive 
precision on the angle $\beta$:
\begin{equation}\label{TOT}
(\sin 2\beta)_{\rm tot}=0.705^{+0.042}_{-0.032}, \qquad 
 \beta=(22.4 \pm 1.5)^\circ~.
\end{equation}
On the other hand $(\sin 2\beta)_{\rm ind}$ obtained by means of 
the five steps only is found to be 
\cite{BUPAST} 
\be
(\sin 2\beta)_{ind}=0.685 \pm 0.052
\label{ind}
\ee
demonstrating an excellent agreement (see also fig.~\ref{fig:figmfv}) 
between the direct measurement in (\ref{ga})
and the standard analysis of
the UT within the SM.
This gives a strong indication that the CKM matrix is very likely 
the dominant source of CP violation in flavour violating decays.
In order to be sure whether this is indeed the case other theoretically
clean quantities have to be measured. In particular the angle $\gamma$ 
that is more sensitive to new physics contributions than $\beta$.
We will return to other processes that are 
useful for the determination of the UT in Sections 5 and 6.

\begin{table}[htb!]
\begin{center}
\caption[]{ \small { Values for different quantities from 
the update of \cite{BUPAST}.
$\lambda_t=V_{ts}^*V_{td}$.}}
\begin{tabular}{|c|c|c|}
\hline
  Strategy       &               UUT              &            SM         \\
  $\bar {\eta}$  &       ~ ~ 0.361 $\pm$ 0.032~~  &   ~~0.341 $\pm$ 0.028~~  \\ 
                 &                                &                        \\
  $\bar {\varrho}$  &       ~~0.149 $\pm$ 0.056~~      &   ~~ 0.178 $\pm$ 0.046~~   \\
                 &                                &                         \\
  $\sin 2\beta$  &  ~~0.715 $^{+0.037}_{-0.034}$~~   & ~~0.705 $^{+0.042}_{-0.032}$~~ \\
                 &                                &                         \\
  $\sin 2\alpha$ &     ~~0.03 $\pm$ 0.31~~         & ~~ --0.19 $\pm$ 0.25~~           \\
                 &                                &                       \\
  $\gamma$       &   ~~$ (67.5 \pm 8.9)^\circ$~~    &   ~~$ (61.5 \pm 7.0)^\circ $~~\\
                 &                                &                       \\
  $R_b$          &       ~~0.393 $\pm$ 0.025~~      & ~~0.390 $\pm$ 0.024~~   \\
                 &                                &                        \\
  $R_t$          &       ~~$0.925\pm 0.060$~~        &    ~~$0.890 \pm 0.048 $~~  \\
                 &                                &                        \\
  $\Delta M_s$ ($ps^{-1}$)   & ~~17.3$^{+2.1}_{-1.3}$~~   &  ~~18.3$^{+1.7}_{-1.5}$~~       \\
                 &                                &                       \\
$\vtd~(10^{-3})$ &        ~~8.61 $\pm$ 0.55~~       &      ~~8.24 $\pm$ 0.41~~       \\
                 &                                &                         \\
~~${\rm Im} \lambda_t$ ($10^{-4}$)~~  & ~~1.39 $\pm$ 0.12~~    &    ~~1.31 $\pm$ 0.10~~    \\
                 &                                &               \\
\hline
\end{tabular}
\label{mfv}
\end{center}
\end{table}

\section{$\epe$ in the Standard Model}\label{EpsilonPrime}
\subsection{Preliminaries}
The ratio $\epe$ that parametrizes the size of direct CP violation with
respect to the indirect CP violation in $K_L\to \pi\pi$ decays has been the 
subject of very intensive experimental and theoretical studies in the last
three decades. After tremendous efforts, on the experimental side the world
average based on the recent results from NA48 \cite{NA48} and KTeV
\cite{KTeV}, and previous results from NA31 \cite{NA31} and E731
\cite{E731}, reads
\begin{equation}
  \label{eps}
  \epe=(16.6\pm 1.6) \cdot 10^{-4} \qquad\qquad (2003)~.
\end{equation}
On the other hand, the theoretical estimates of this ratio are subject to
very large hadronic uncertainties. While several analyzes of recent years
within the Standard Model (SM) find results that are compatible with 
(\ref{eps}) \cite{EP99,BRMSSM,ROME,DORT,Trieste,PP,Lund,MARS}), 
it is fair to say
that the chapter on the theoretical calculations of
$\epe$ is far from being closed. A full historical account of the theoretical
efforts before 1998 can for example be found in \cite{AJBLH,BFE00}.
See also \cite{Bert02}.

It should be emphasized that all existing analyzes of $\epe$
 use the NLO Wilson coefficients calculated by the Munich and Rome 
groups in 1993 \cite{BJLW92,BJL93,CFMR93}
 but the hadronic matrix elements, the main theoretical uncertainty in $\epe$,
  vary from paper to paper. 
Nevertheless, apart from the hadronic matrix element of the dominant QCD
penguin operator $Q_6$, in the last years progress has been made with the
determination of all other relevant parameters, which enter the theoretical
prediction of $\epe$. Let me review then briefly the present situation. 
Further details can be found in \cite{BJ03}.

\subsection{Basic Formulae}
The central formula for $\epe$ of \cite{EP99,BRMSSM,BJL93,BJ03} 
can be cast into the following approximate expression (it reproduces the 
results in \cite{BJ03} to better than $2\%$)
\be \frac{\varepsilon'}{\varepsilon}= \IM\lambda_t
\cdot F_{\varepsilon'}(x_t), \qquad \lambda_t=V^*_{ts}V_{td},
\label{epeth}
\ee
\be
F_{\varepsilon'}(x_t) =\left[18.7~R_6(1-\Omega_{\rm IB})-6.9~R_8-1.8\right]
\left[\frac{\Lms^{(4)}}{340\mev}\right]
\label{FE}
\ee
with
the non-perturbative parameters $R_6$ and $R_8$ 
defined as
\be\label{RS}
R_6\equiv \bsi\left[ \frac{121\mev}{m_s(m_c)+m_d(m_c)} \right]^2,
\qquad
R_8\equiv \frac{\bei}{\bsi} R_6.
\ee
The hadronic $B$-parameters $\bsi$ and $\bei$ represent
the matrix
elements of the dominant QCD-penguin ($Q_6$) and the dominant electroweak
penguin ($Q_8$) operator. 
In the large--$N_c$ approach of \cite{BBG80} they are 
given by \cite{BJL93,BBH90}
\begin{eqnarray}\label{LARGEN}
\langle Q_6 \rangle_0~&=&\!\! -\,4\,\sqrt{\frac{3}{2}}\,(F_K-F_\pi)
\biggl(\!\frac{m_K^2}{121\mev}\!\biggr)^{\!\!2} R_6 \,=\,
-\,0.597\cdot R_6\,\gev^3 \,, \\
\langle Q_8 \rangle_2~ &=& \sqrt{3}\,F_\pi
\biggl(\!\frac{m_K^2}{121\mev}\!\biggr)^{\!\!2} R_8 \,=\,
0.948\cdot R_8\,\gev^3 \,.
\end{eqnarray}
Finally $\Omega_{\rm IB}=0.06\pm0.08$ \cite{CPEN} represents isospin 
breaking correction.

In the strict large--$N_c$ limit, $B_6^{(1/2)}=B_8^{(3/2)}=1$ and 
\be\label{RELP}
\frac{R_6}{R_8}=1~, \qquad
\frac{\langle Q_6 \rangle_0}{\langle Q_8 \rangle_2}=-\,0.63~,
\ee
 so that there 
is a one-to-one correspondence
between $\epe$ and $R_6=R_8$ for fixed values of the remaining parameters.
Moreover, only for certain values of $m_s(m_c)$ is one able
to obtain the experimental value for $\epe$ \cite{KNS99}.
 Note that once $m_s(m_c)$ is known, also $R_6$, $R_8$, $\langle Q_6 \rangle_0$
and $\langle Q_8 \rangle_2$ are known, but they always satisfy the relations
in (\ref{RELP}).

\subsection{Numerical Results}
The relevant input parameters are as follows. First
\begin{equation}\label{ltms}
\IM\lambda_t = (1.31 \pm 0.10)\cdot 10^{-4}, \qquad
m_s(m_c) = 115 \pm 20 \,\mev~
\end{equation}
where
the value of $m_s$ is
an average over recent determinations
(see references in \cite{BJ03}). The central value  corresponds to
$m_s(2\,\gev)=100\,\mev$.

Concerning $\langle Q_8 \rangle_2$, in the last
years progress has been achieved both in the framework of lattice QCD
as well as with analytic methods \cite{Lund,MARS,Lattepe,Bec02,CDGM03,Nar00}.
The current status of $\langle Q_8\rangle_2$  has
been summarized nicely in \cite{CDGM03}. The
most precise determination of $\langle Q_8\rangle_2$ comes from 
the lattice QCD measurement \cite{Bec02}, corresponding to 
$R_8=0.81 \pm 0.08$. Several analytic methods give higher results but
compatible with it.  
We will use \cite{BJ03}
\be
\label{R8}
R_8=0.8 \pm 0.2~, \qquad
\langle Q_8\rangle_2^{{\rm NDR}}(m_c) = (0.76 \pm 0.19)\,\gev^3~.
\ee

The situation is less clear concerning $\langle Q_6\rangle_0$ but 
assuming that new physics contributions to $\epe$ 
can be neglected one finds from (\ref{eps})--(\ref{FE}) and (\ref{R8}) 
that \cite{BJ03} 
\begin{equation}
\label{R6}
R_6 = 1.15\pm 0.16~, \qquad 
\langle Q_6\rangle_0^{{\rm NDR}}(m_c) = - (0.69 \pm 0.10)\,\gev^3~.
\end{equation}

\begin{figure}[htb]
\begin{center}
\includegraphics[angle=270, width=11cm]{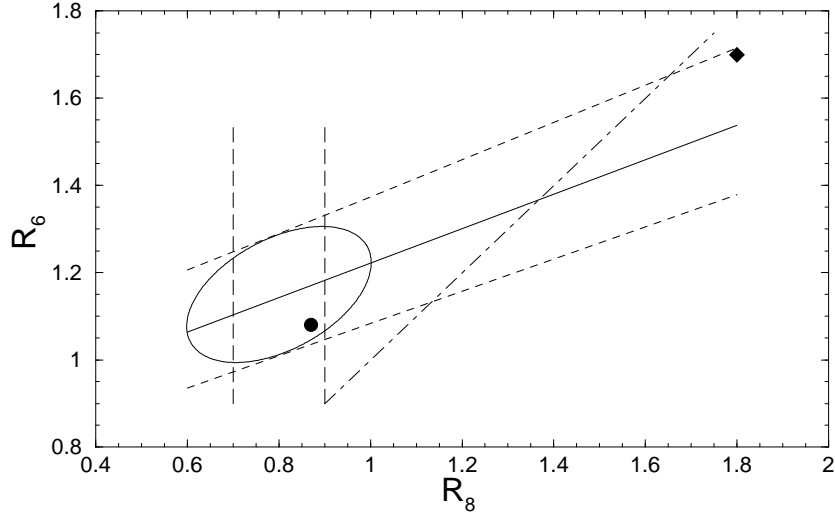}
\end{center}
\caption{$R_6$ as a function of $R_8$. For a detailed explanation see
the text.
\label{fig1}}
\end{figure}
More generally, the correlation between $R_8$ and $R_6$ that 
is implied by the data on $\epe$ is shown  in the spirit
of the ``$\epe$--path" of \cite{BG01} (see also \cite{Don01,MARS})
in figure~\ref{fig1} \cite{BJ03}. The solid straight line corresponds to the
central values of parameters, whereas the short-dashed lines are
the uncertainties due to a variation of the input parameters. 
The
vertical long-dashed lines indicate the lattice range for $R_8$ \cite{Bec02},
whereas the ellipse describes the correlation between $R_6$ and $R_8$ implied
by the data on $\epe$ when taking into account the more conservative constraint
on $R_8$ given in (\ref{R8}). The value in (\ref{R6}) corresponds to this 
ellipse.
The full circle and diamond in figure~\ref{fig1} represent 
the central results
of \cite{PP} and \cite{MARS} respectively, that are discussed in 
detail in \cite{BJ03}, and the dashed-dotted line shows the strict large-$N_c$
relation $R_6=R_8$.

In table~\ref{tab:eps} we show $\epe$ for specific 
values of $R_6$, $R_8$ and $\Lms^{(4)}$, and 
 in table~\ref{tab:LN} 
 as a function of $m_s(m_c)$ and $\Lms^{(4)}$ obtained 
in the strict large--$N_c$ limit. Here ${\rm Im}\lambda_t=1.34\cdot 10^{-4}$.

\begin{table}[htb]
\caption[]{ $\epe$ in units of $10^{-4}$ for $m_t = 165 \gev$ 
and various $\Lms^{(4)}$, $R_6$, $R_8$. \label{tab:eps}}
\begin{center}
\begin{tabular}{|c||c|c||c|c||c|c|}
\hline
& \multicolumn{2}{c||}{$R_6 = 1.00$} &
  \multicolumn{2}{c||}{$R_6 = 1.15$} &
  \multicolumn{2}{c| }{$R_6 = 1.30$} \\
\hline
$\Lms^{(4)}~[\mev] $ & 
$R_8=0.8$ & $R_8=1.0$ &
$R_8=0.8$ & $R_8=1.0$ &
$R_8=0.8$ & $R_8=1.0$  \\
\hline
310 & 12.2 &  10.5 & 15.4 & 13.7 & 18.6 & 16.9 \\
340 & 13.3 &  11.5 & 16.8 & 14.9 & 20.2 & 18.4 \\
370 & 14.6 & 12.6 & 18.3 & 16.3 & 22.0 & 20.0 \\
\hline
\end{tabular}
\end{center}
\end{table}

\begin{table}[htb]
\caption[]{The ratio $\epe$ in units of $10^{-4}$ for the strict large--$N_c$
results $\bsi=\bei=1.0$, $\hat B_K=0.75$ and various values of $\Lms^{(4)}$
and $m_s(m_c)$. 
\label{tab:LN}}
\begin{center}
\begin{tabular}{|c||c|c|c|}
\hline
$\Lms^{(4)}~[\mev] $ & $m_s(m_c)=115\mev$ & $m_s(m_c)=105\mev$ &
$m_s(m_c)=95\mev$  \\
\hline
310 &  10.8 & 13.3 & 16.5  \\
340 & 11.8 & 14.4 & 18.0  \\
370 & 12.9 & 15.8 & 19.6 \\
\hline
\end{tabular}
\end{center}
\end{table}

\subsection{Conclusions}
There are essentially two messages from this analysis \cite{BJ03}:
\begin{itemize}
\item
If indeed $R_8=0.8\pm 0.2$ as indicated by several recent estimates, then
the data on $\epe$ imply
\begin{equation}
\label{Q6Q8a}
R_6 = 1.15 \pm 0.16~, \quad
\frac{R_6}{R_8}\approx 1.4~, \quad 
\langle Q_6\rangle_0^{{\rm NDR}}(m_c) \approx 
-\,\langle Q_8\rangle_2^{{\rm NDR}}(m_c)~.
\end{equation}
This is in accordance with the results in \cite{PP}, but differs 
from the large--$N_c$ approach in \cite{BBG80} 
in which $R_6\approx R_8$ and
$\langle Q_6\rangle_0^{{\rm NDR}}(m_c)$ is chirally suppressed relatively to 
$\langle Q_8\rangle_2^{{\rm NDR}}(m_c)$.
\item
The large--$N_c$ approach of \cite{BBG80} can only be made
consistent with data provided 
\be
R_6=R_8=1.36\pm 0.30
\ee
and $\langle Q_8\rangle_2^{{\rm NDR}}(m_c)$ is  higher
than obtained by most recent approaches reviewed in \cite{BJ03,CDGM03}.
This requires $m_s(m_c)\le 105\mev$ which is on the low side of
(\ref{ltms}) but close to low values of $m_s(m_c)$  indicated by the
most recent lattice simulations with dynamical fermions \cite{HCLMT02,Gup03}. 
\end{itemize}

Large non-factorizable contributions to   
$\langle Q_6\rangle_0$ and $\langle Q_8\rangle_2$ are found in the 
chiral limit in the large N 
approach of \cite{MARS} and consequently
the structure of the matrix 
elements in question differs in this approach from the formulae
(\ref{LARGEN}). Interestingly, in spite of these large non-factorizable 
contributions, the relation $R_6=R_8$ is roughly satisfied in this approach
(see fig.~\ref{fig1}) but one has to go beyond the chiral limit  
to draw definite conclusions.

As seen  in figure~\ref{fig1}, all these three scenarios are consistent 
with the data. Which of these pictures of $\epe$ is correct, can only be 
answered by calculating $\langle Q_6\rangle_0$, $\langle Q_8\rangle_2$
and $m_s$ accurately 
by means of non-perturbative 
methods that are reliable. Such calculations are independent of the 
assumption about
the role of new physics in $\epe$ that has been made in \cite{BJ03} 
in order to extract
$\langle Q_6\rangle_0$ from the data.
If the values for  $R_{6,8}$ will be found one day 
to lie significantly outside the allowed region in figure~\ref{fig1}, new
physics contributions to $\epe$ will be required in order to fit the
experimental data.

\section{ The Angles \boldmath{$\alpha$}, \boldmath{$\beta$} and 
\boldmath{$\gamma$} from B Decays}
\subsection{Preliminaries}

CP violation in B decays is certainly one of the most important 
targets of B-factories and of dedicated B-experiments at hadron 
facilities. It is well known that CP-violating effects are expected
to occur in a large number of channels at a level attainable 
experimentally in the near future.
Moreover there exist channels which
offer the determination of CKM phases essentially without any hadronic
uncertainties. 

The first results on
$\sin 2\beta$ from BaBar and Belle are 
very encouraging. These results should be further improved over the coming 
years 
through the new measurements of $a_{\psi K_S}(t)$ by both collaborations 
and by CDF and D0 at Fermilab. Moreover measurements of CP 
asymmetries in other B decays and the measurements of the angles 
$\alpha,~\beta$ and $\gamma$ by means of various strategies using
two-body B decays should contribute substantially to our understanding
of CP violation and will test the KM picture of CP violation.

The various types of CP violation have been already classified in Section 2.
It turned out that CP violation in the interference 
of mixing and decay, in a B meson decay into a CP eigenstate, is very
suitable for a theoretically clean determination of the angles of the 
unitarity triangle provided a single CKM phase governs  the decay. 
However as we will see below several useful strategies 
for the determination of the angles $\alpha,~\beta$ and $\gamma$
have been developed that are effective also in the presence of competing 
CKM phases and when the final state in not a CP eigenstate. 
The notes below should only be considered as an introduction to this 
reach field. For more details the references in Section 1 should be 
contacted.

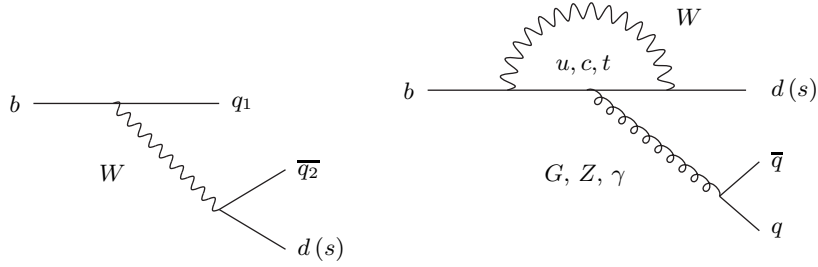
\begin{figure}[t]
\vspace*{0.2truecm}
{\small
\hspace*{3.2truecm}\begin{picture}(80,50)(80,20)
\Line(10,45)(80,45)\Photon(40,45)(80,5){2}{10}
\Line(80,5)(105,20)\Line(80,5)(105,-10)
\Text(5,45)[r]{$b$}\Text(85,45)[l]{$q_1$}
\Text(110,20)[l]{$\overline{q_2}$}
\Text(110,-10)[l]{$d\,(s)$}
\Text(45,22)[tr]{$W$}
\end{picture}}
\hspace*{-0.6truecm}
{\small
\begin{picture}(140,60)(0,20)
\Line(10,50)(130,50)\Text(5,50)[r]{$b$}\Text(140,50)[l]{$d\,(s)$}
\PhotonArc(70,50)(30,0,180){3}{15}
\Text(69,56)[b]{$u,c,t$}\Text(109,75)[b]{$W$}
\Gluon(70,50)(120,10){2}{10}
\Line(120,10)(135,23)\Line(120,10)(135,-3)
\Text(85,22)[tr]{$G$, $Z$, $\gamma$}\Text(140,-3)[l]{$q$}
\Text(140,23)[l]{$\overline{q}$}
\end{picture}}
\vspace*{1.2truecm}
\caption{Tree and penguin diagrams.}\label{TP}
\end{figure}

\subsection{Classification of Elementary Processes}
Non-leptonic B decays are caused by elementary decays of b quarks that
are represented by tree and penguin diagrams in fig.~\ref{TP}. Generally we 
have
\be
b\to q_1 \bar q_2 d(s), \qquad  b \to q\bar q d(s)
\ee
for tree and penguin diagrams, respectively.

There are twelve basic transitions that can be divided into three classes:

{\bf Class I}: both tree and penguin diagrams contribute. Here 
$q_1=q_2=q=u,c$ and consequently the basic transitions are
\be
b\to c\bar c s, \qquad b \to c\bar c d, \qquad
b\to u\bar u s, \qquad b \to u\bar u d.
\ee

{\bf Class II}: only tree diagrams contribute. Here 
$q_1\not=q_2\in\{u,c\}$ and 
\be
b\to c\bar u s, \qquad b \to c\bar u d, \qquad
b\to u\bar c s, \qquad b \to u\bar c d. 
\ee

{\bf Class III}: only penguin diagrams contribute. Here 
$q=d,s$ and 
\be
b\to s\bar s s, \qquad b \to s\bar s d, \qquad
b\to d\bar d s, \qquad b \to d\bar d d. 
\ee

Now in presenting various decays below, we did not show the corresponding 
diagrams on purpose. Afterall these are lectures and the exercise for the 
students is to draw these diagrams by embedding the elementary diagrams 
of fig.~\ref{TP}  into a given B meson decay.  
In case of difficulties the student 
should look at \cite{LHCB,REV} where these diagrams can be found.

\subsection{Neutral B Decays into CP eigenstates}
\subsubsection{\boldmath{$B^0_d\to J/\psi K_S$} and 
\boldmath{$\beta$}}
The amplitude for this decay can be written as follows
\be\label{FA1}
A(B^0_d\to J/\psi K_S)=V_{cs}V_{cb}^*(A_T+P_c)+ V_{us}V_{ub}^*P_u+
V_{ts}V_{tb}^*P_t
\ee
where $A_T$ denotes tree diagram contributions and $P_i$ with $i=u,c,t$ 
stand for penguin diagram contributions with internal $u$, $c$ and $t$ 
quarks. Now
\be
V_{cs}V_{cb}^*\approx A\lambda^2, \quad
V_{us}V_{ub}^*\approx A \lambda^4 R_b e^{i\gamma}, 
\quad V_{ts}V_{tb}^*=- V_{us}V_{ub}^*- V_{cs}V_{cb}^*
\ee
with the last relation following from the unitarity of the CKM matrix.
Thus
\be\label{FA1a}
A(B^0_d\to J/\psi K_S)=V_{cs}V_{cb}^*(A_T+P_c-P_t)+ V_{us}V_{ub}^*(P_u-P_t)~.
\ee
We next note that
\be
\left|\frac{V_{us}V_{ub}^*}{V_{cs}V_{cb}^*}\right|\le 0.02, \qquad
\frac{P_u-P_t}{A_T+P_c-P_t}\ll 1
\ee
where the last inequality is very plausible as the Wilson coefficients 
of the current--current operators responsible for $A_T$ are much larger 
than the ones of the penguin operators \cite{AJBLH,BBL}. 
Consequently this decay is dominated 
by a single CKM factor and as discussed in Section 2, a clean determination 
of the relevant CKM phase is possible. Indeed in this decay $\phi_D=0$  
and $\phi_M=-\beta$. Using (\ref{simple})
we find then ($\eta_{J/\psi K_S}=-1$)
\begin{equation}\label{psiKS}
a^{\rm int}_{CP}(J/\psi K_S)=
\eta_{J/\psi K_S}\sin(2\phi_D-2\phi_M)=-\sin 2\beta,
\end{equation}
\be
C_{J/\psi K_S}=0, \qquad S_{J/\psi K_S}=\sin 2\beta
\ee
that is confirmed by experiment as discussed in Section 3.
\subsubsection{\boldmath{$B^0_s\to J/\psi \phi$} and 
\boldmath{$\beta_s$}}
This decay differs from the previous one by the spectator quark, with 
$d\to s$  so that the formulae above remain unchanged except that 
now $\phi_M=-\beta_s=-\lambda^2\bar\eta$. A complication arises as 
the final state is an admixture of $CP=+$ and $CP=-$ states. This issue 
can be resolved experimentally \cite{LHCB}. 
Choosing $\eta_{J/\psi \phi}=1$ we then find
\begin{equation}\label{psiphi}
a^{\rm int}_{CP}(J/\psi \phi)=
\sin(2\phi_D-2\phi_M)=2\beta_s=2\lambda^2\bar\eta\approx 0.03,
 \qquad C_{J/\psi \phi}=0~.
\end{equation}
Thus this asymmetry measures the phase of $V_{ts}$ that is predicted to 
be very small from the unitarity of the CKM matrix alone. Because of this
there is a lot of room for 
new physics contributions here. 
\subsubsection{\boldmath{$B^0_d\to \phi K_S$} and 
\boldmath{$\beta$}}
This decay proceeds entirely through penguin diagrams and consequently 
should be much more sensitive to new physics contributions than the decay
$B^0_d\to J/\psi K_S$. Assuming $\phi=(s\bar s)$, the decay amplitude 
is given by (\ref{FA1a}) 
with $A_T$ removed:
\be\label{FA1b}
A(B^0_d\to \phi K_S)=V_{cs}V_{cb}^*(P_c-P_t)+ V_{us}V_{ub}^*(P_u-P_t)~.
\ee
With 
\be
\left|\frac{V_{us}V_{ub}^*}{V_{cs}V_{cb}^*}\right|\le 0.02, \qquad
\frac{P_u-P_t}{P_c-P_t}=\ord( 1)
\ee
also in this decay single CKM phase dominates and as 
$\phi_D$ and $\phi_M$ are the same as in $B^0_d\to J/ \psi K_S$ we 
find 
\be\label{phi K_S}
C_{\phi K_S}=0, \qquad S_{\phi K_S}= S_{J/\psi K_S}=\sin 2\beta~.
\ee
The equality of these two asymmetries need not be perfect as the $\phi$ 
meson is not entirely a $s\bar s$ state and the approximation of neglecting 
the second amplitude in (\ref{FA1b}) could be only true within a few percent.
However, a detailed analysis shows \cite{Worah} that these two 
asymmetries should 
be very close to each other within the SM:  
$|S_{\phi K_S}-S_{J/\psi K_S}|\le 0.04$~. Any strong violation of this 
bound would be a signal for new physics.

In view of this prediction, the first results on this asymmetry from 
BaBar \cite{Bab} and Belle \cite{Bela} are truely exciting:
\begin{displaymath}\label{sinbphi}
(\sin 2\beta)_{\phi K_S}=\left\{
\begin{array}{ll}
-0.19 \pm 0.51 \, \mbox{(stat)} \pm0.09 \, \mbox{(syst)} & \mbox{(BaBar)}\\
-0.73\pm 0.64 \, \mbox{(stat)} \pm0.18  \, \mbox{(syst)} & \mbox{(Belle).}
\end{array}
\right.
\end{displaymath}
Consequently
\be\label{gaphi}
S_{\phi K_s}=-0.39\pm 0.41, \qquad C_{\phi K_s}=0.56\pm 0.43, 
\ee
\be
|S_{\phi K_S}-S_{J/\psi K_S}|=1.12\pm 0.41
\ee
where the result for $C_{\phi K_S}$, that is consistent with zero, comes 
solely from Belle. We observe that the bound 
$|S_{\phi K_S}-S_{J/\psi K_S}|\le 0.04$ is violated by $2.7\sigma$. While 
this is still insufficient to claim the presence of new physics, the 
fact that the two asymmetries are found to be quite different, invited 
a number of theorists to speculate what kind of new physics could be 
responsible for this difference. 
Some references are given in \cite{PHIKS}.
Enhanced QCD penguins, enhanced $Z^0$ 
penguins, rather involved supersymmetric scenarios have been suggested as 
possible origins of the departure from the SM prediction. I have no space 
to review these papers and although I find a few of them quite interesting, 
it is probably better to wait until the experimental errors decrease.

Of interest are also the measurements 
\begin{displaymath}\label{sinbeta}
S_{\eta' K_S}=\left\{
\begin{array}{ll}
0.02 \pm 0.35  & \mbox{(BaBar)}\\
0.76\pm 0.36 & \mbox{(Belle).}
\end{array}
\right.
\end{displaymath}
and 
$C_{\eta' K_S}=-0.26\pm 0.22$ from Belle that are fully consistent with 
$(\sin 2\beta)_{J/\psi K_S}$. At first sight one could wonder why this
asymmetry differs from $(\sin 2\beta)_{\phi K_S}$ as the decay in question 
is also penguin dominated and $\eta'\approx (s\bar s)$, but the fact that 
$\eta'$ deviates from a pure $(s\bar s)$ state more than $\phi$ allows 
for some small contributions involving tree diagrams that could spoil 
the exact equality of these two asymmetries. 
\subsubsection{\boldmath{$B^0_d\to \pi^+\pi^-$} and 
\boldmath{$\alpha$}}
This decay receives the contributions from both tree and penguin diagrams.
The amplitude  can be written as follows
\be\label{FALPHA}
A(B^0_d\to \pi^+\pi^-)=V_{ud}V_{ub}^*(A_T+P_u)+ V_{cd}V_{cb}^*P_c+
V_{td}V_{tb}^*P_t
\ee
where 
\be
V_{cd}V_{cb}^*\approx A\lambda^3, \quad
V_{ud}V_{ub}^*\approx A \lambda^3 R_b e^{i\gamma}, 
\quad V_{td}V_{tb}^*=- V_{ud}V_{ub}^*- V_{cd}V_{cb}^*~.
\ee
Consequently
\be\label{FALPHAa}
A(B^0_d\to \pi^+\pi^-)=V_{ud}V_{ub}^*(A_T+P_u-P_t)+ V_{cd}V_{cb}^*(P_c-P_t).
\ee
We next note that
\be
\left|\frac{V_{cd}V_{cb}^*}{V_{ud}V_{ub}^*}\right|=\frac{1}{R_b}\approx 2.5,
 \qquad
\frac{P_c-P_t}{A_T+P_u-P_t}\equiv \frac{P_{\pi\pi}}{T_{\pi\pi}}.
\ee
Now the dominance of a single CKM amplitude in contrast to the cases 
considered until now is very uncertain and takes only place provided
$P_{\pi\pi}\ll T_{\pi\pi}$. Let us assume first that this is indeed the 
case. Then this decay is dominated 
by a single CKM factor and a clean determination 
of the relevant CKM phase is possible. Indeed in this decay $\phi_D=\gamma$  
and $\phi_M=-\beta$. Using (\ref{simple})
we find then ($\eta_{\pi\pi}=1$)
\begin{equation}\label{pipi}
a^{\rm int}_{CP}(\pi\pi)=
\eta_{\pi\pi}\sin(2\phi_D-2\phi_M)=\sin 2(\gamma+\beta)=-\sin 2\alpha
\end{equation}
and 
\be
C_{\pi\pi}=0, \qquad S_{\pi\pi}=\sin 2\alpha~.
\ee
This should be compared with the first results from BaBar and Belle:
\begin{displaymath}\label{sinaC}
C_{\pi\pi}=\left\{
\begin{array}{ll}
-0.30 \pm 0.25 \, \mbox{(stat)} \pm0.04 \, \mbox{(syst)} & \mbox{(BaBar)}\\
-0.77\pm 0.27 \, \mbox{(stat)} \pm0.08  \, \mbox{(syst)} & \mbox{(Belle)}
\end{array}
\right.
\end{displaymath}

\begin{displaymath}\label{sinaS}
S_{\pi\pi}=\left\{
\begin{array}{ll}
-0.02 \pm 0.34 \, \mbox{(stat)} \pm0.05 \, \mbox{(syst)} & \mbox{(BaBar)}\\
-1.23\pm 0.41 \, \mbox{(stat)} \pm0.08  \, \mbox{(syst)} & \mbox{(Belle).}
\end{array}
\right.
\end{displaymath}
The results from BaBar are consistent with our expectations. Afterall 
$\alpha$ from the UT fit is in the ballpark of $90^\circ$.
On the other hand Belle results indicate a non-zero asymmetry and a 
sizable contribution of the penguin diagrams invalidating our assumption
$P_{\pi\pi}\ll T_{\pi\pi}$. Yet, as the results from BaBar and Belle are
incompatible with each other, the present picture of this decay is not
conclusive and one has to wait for better data. 

The ``QCD penguin pollution" discussed above  has to be
taken care of in order to extract $\alpha$. 
The well known strategy to deal with this "penguin problem''
is the isospin analysis of Gronau and London \cite{CPASYM}. It
requires however the measurement of $Br(B^0\to \pi^0\pi^0)$ which is
expected to be below $10^{-6}$: a very difficult experimental task.
For this reason several, rather involved, strategies  
have been proposed which
avoid the use of $B_d \to \pi^0\pi^0$ in conjunction with
$a_{CP}(\pi^+\pi^-,t)$. They are reviewed in 
\cite{BF97,BABAR,LHCB,REV}. 
The most recent analyses of $B\to\pi\pi$, also related to the determination 
of $\gamma$ and $(\bar\varrho,\bar\eta)$, can be found in 
\cite{CERNCKM,ALPHA,FLISMA}.

While I have some doubts that a precise value of $\alpha$ will 
follow in a foreseable future from this enterprise,
one should also stress \cite{BBSIN,BBNS2,BUPAST} that only a 
moderately precise measurement of $\sin 2\alpha$ can be as useful for 
the UT as a precise measurement of the angle $\beta$.  
This is clear from table~\ref{mfv} that
shows very large uncertainties in the indirect determination of 
$\sin 2\alpha$.

\subsection{Decays to CP Non-Eigenstates}
\subsubsection{Preliminaries}
The strategies discussed below have the following general properties:
\begin{itemize}
\item
$B^0_d(B^0_s)$ and their antiparticles $\bar B^0_d(\bar B^0_s)$ can decay to 
the same final state,
\item
Only tree diagrams contribute to the decay amplitudes,
\item
A full time dependent analysis of the four processes is required:
\be
B^0_{d,s}(t)\to f, \quad \bar B^0_{d,s}(t)\to f, \quad
 B^0_{d,s}(t)\to \bar f, \quad \bar B^0_{d,s}(t)\to \bar f~.
\ee
\end{itemize}
The latter analysis allows to measure
\begin{equation}\label{4rates}
\xi_f=\exp(i2\phi_M)\frac{A(\bar B^0\to f)}{A(B^0 \to f)}, \qquad
\xi_{\bar f}=\exp(i2\phi_M)\frac{A(\bar B^0\to \bar f)}{A(B^0 \to \bar f)}.
\end{equation}
It turns out then that
\be
\xi_f\cdot \xi_{\bar f}=F(\gamma,\phi_M)
\ee
without any hadronic uncertainties, so that determining $\phi_M$ from 
other decays as discussed above, allows the determination of $\gamma$.
Let us show this.
\subsubsection{\boldmath{$B^0_d\to D^\pm\pi^\mp$}, 
\boldmath{$\bar B^0_d\to D^\pm\pi^\mp$} and \boldmath{$\gamma$} }
With $f=D^+\pi^-$ the four decay amplitudes are given by
\be\label{amp12}
A(B^0_d\to D^+\pi^-)=M_f A \lambda^4 R_b e^{i\gamma}, \qquad
A(\bar B^0_d\to D^+\pi^-)= \bar M_f A \lambda^2
\ee
\be\label{amp34}
A(\bar B^0_d\to D^-\pi^+)=\bar M_{\bar f} A \lambda^4 R_b e^{-i\gamma}, \qquad
A(B^0_d\to D^-\pi^+)=  M_{\bar f} A \lambda^2
\ee
where we have factored out the CKM parameters, $A$ is a Wolfenstein 
paramater and $M_i$ stand for the rest of the amplitudes that generally are 
subject to large hadronic uncertainties. The important point is that each 
of these transitions receives the contribution from a single phase so that
\begin{equation}\label{2xiD}
\xi_f^{(d)}=
e^{-i(2\beta+\gamma)}\frac{1}{\lambda^2 R_b}\frac{\bar M_f}{M_f}, \qquad
\xi_{\bar f}^{(d)}=
e^{-i(2\beta+\gamma)}\lambda^2 R_b\frac{\bar M_{\bar f}}{M_{\bar f}}~.
\end{equation}
Now, as CP is conserved in QCD we simply have
\be
M_f=\bar M_{\bar f}, \qquad \bar M_f= M_{\bar f} 
\ee
and consequently \cite{DPI}
\be\label{dxi}
\xi_f^{(d)}\cdot \xi_{\bar f}^{(d)}=e^{-i2(2\beta+\gamma)}
\ee
as promised. The phase $\beta$ is already known with high precision 
and consequently $\gamma$ can be determined. Unfortunately as seen in
(\ref{amp12}) and (\ref{amp34}), the relevant interefences are 
$\ord(\lambda^2)$ and the execution of this strategy is a very difficult 
experimental task. See \cite{Silva} for an interesting discussion.

\subsubsection{\boldmath{$B^0_s\to D_s^\pm K^\mp$}, 
\boldmath{$\bar B^0_s\to D_s^\pm K^\mp$} and \boldmath{$\gamma$}}
Replacing the d-quark by the s-quark in the strategy just discussed allows 
to solve the latter problem. With $f=D^+_s K^-$ equations 
(\ref{amp12}) and (\ref{amp34}) are replaced by
\be\label{samp12}
A(B^0_s\to D_s^+K^-)=M_f A \lambda^3 R_b e^{i\gamma}, \qquad
A(\bar B^0_s\to D_s^+K^-)= \bar M_f A \lambda^3
\ee
\be\label{samp34}
A(\bar B^0_s\to D_s^- K^+)=\bar M_{\bar f} A \lambda^3 R_b e^{-i\gamma}, \qquad
A(B^0_s\to D_s^- K^+)=  M_{\bar f} A \lambda^3~.
\ee
Proceeding as in the previous strategy one finds \cite{adk}
\be\label{sxi}
\xi_f^{(s)}\cdot \xi_{\bar f}^{(s)}=e^{-i2(2\beta_s+\gamma)}
\ee
with $\xi_f^{(s)}$ and $\xi_{\bar f}^{(s)}$ being the analogs of 
$\xi_f^{(d)}$ and $\xi_{\bar f}^{(d)}$, respectively. Now, all interferring 
amplitudes are of a similar size. With $\beta_s$ extracted 
one day from the asymmetry in $B_s^0(\bar B_s^0)\to J/ \psi\phi$, the angle 
$\gamma$ can be determined.
\subsubsection{\boldmath{$B^\pm\to D^0 K^\pm$}, 
\boldmath{$B^\pm\to \bar D^0 K^\pm$} and \boldmath{$\gamma$} }
By replacing the spectator s-quark in the last strategy through the 
u-quark one arrives at decays of $B^\pm$ that can be used to extract 
$\gamma$. Also this strategy is unaffected by penguin contributions. 
Moreover, as particle-antiparticle mixing is absent here, $\gamma$ can 
be measured directly without any need for phases in the mixing. Both 
these features make it plausible that this strategy, not involving 
to first approximation any loop diagrams,  is particularly suited 
for the determination of $\gamma$ without any new physics pollution.

By considering six decay rates $B^{\pm}\to D^0_{CP} K^{\pm}$,
$B^+ \to D^0 K^+,~ \bar D^0 K^+$ and  $B^- \to D^0 K^-,~ \bar D^0 K^-$
where $D^0_{CP}=(D^0+\bar D^0)/\sqrt{2})$ is a CP eigenstate, and 
noting that
\be\label{CONST}
A(B^+ \to \bar D^0 K^+)= A(B^- \to D^0 K^-),
\ee
\be\label{CONSTa}
A(B^+ \to D^0 K^+)= A(B^- \to \bar D^0 K^-) e^{2 i\gamma}
\ee
the well known triangle construction due to Gronau and Wyler 
\cite{Wyler} allows to determine $\gamma$. However, the method is 
not without problems. The detection of $D^0_{CP}$, that is necessary 
for this determination because $K^+\bar D^0\not=K^+ D^0$, is experimentally 
challenging.
Moreover, the small
branching ratios of the colour supressed channels in 
(\ref{CONSTa}) and the absence of this suppression in the two
remaining channels in (\ref{CONST}) imply a rather 
squashed triangle 
thereby making the extraction of $\gamma$ very difficult.
Variants of this method
that could be more promising are discussed in \cite{DUN2,V97}.
\subsubsection{Other Clean Strategies for \boldmath{$\gamma$} and 
\boldmath{$\beta$}}
The three strategies discussed above can be generalized to other decays. 
In particular \cite{DUN2,FLEISCHER}
\begin{itemize}
\item
$2\beta+\gamma$ and $\gamma$ can be measured in
\be
B^0_d\to K_S D^0,~K_S \bar D^0, \qquad B^0_d\to \pi^0 D^0,~\pi^0 \bar D^0 
\ee
and the corresponding CP conjugated channels,
\item
$2\beta_s+\gamma$ and $\gamma$ can be measured in
\be
B^0_s\to \phi D^0,~\phi \bar D^0, \qquad B^0_s\to K_S^0 D^0,~K_S \bar D^0 
\ee
and the corresponding CP conjugated channels,
\item
$\gamma$ can be measured by generalizing the Gronau--Wyler construction 
to $B^\pm\to D^0\pi^\pm, \bar D^0\pi^\pm$ and to $B_c$ decays \cite{FW01}:
\be
B_c^\pm\to D^0 D^\pm_s,~\bar D^0 D^\pm_s, 
\qquad B_c^\pm\to D^0 D^\pm,~\bar D^0 D^\pm~. 
\ee
\end{itemize}
In this context I can strongly recommend recent papers by Fleischer 
\cite{FLEISCHER} 
that while discussing these decays go far beyond the methods presented
here. It appears that the methods discussed in this subsection  may give 
useful results at later stages of CP-B investigations, in particular 
at LHC-B and BTeV. 
\subsection{U--Spin Strategies}
\subsubsection{Preliminaries}
 Useful strategies for $\gamma$ using the U-spin symmetry have
been proposed in \cite{RF99,RF991}. The first strategy involves
the decays $B^0_{d,s}\to \psi K_S$ and $B^0_{d,s}\to D^+_{d,s} D^-_{d,s}$.
The second strategy involves $B^0_s\to K^+ K^-$ and $B^0_d\to\pi^+\pi^-$.
They are unaffected by FSI and are only limited
by U-spin breaking effects. They are promising for
Run II at FNAL and in particular for LHC-B. 

A method of determining $\gamma$, using $B^+\to K^0\pi^+$ and the
U-spin related processes $B_d^0\to K^+\pi^-$ and $B^0_s\to \pi^+K^-$,
was presented in \cite{GRCW}. A general discussion of U-spin symmetry 
in charmless B decays and more references to this topic can be
found in \cite{REV,G00}. I will only briefly discuss the 
method in \cite{RF991}.
\subsubsection{\boldmath{$B^0_d\to \pi^+\pi^-$},  
\boldmath{$B^0_s\to K^+K^-$} and \boldmath{$(\gamma,\beta)$}}
Replacing in $B^0_d\to \pi^+\pi^-$ the $d$ quark by the $s$ quark we obtain 
the decay $B^0_s\to K^+K^-$. 
The amplitude  can be then written in analogy to (\ref{FALPHAa}) as follows
\be\label{Fgamma}
A(B^0_s\to K^+K^-)=V_{us}V_{ub}^*(A'_T+P'_u-P'_t)+ V_{cs}V_{cb}^*(P'_c-P'_t).
\ee
This formula differs from (\ref{FALPHAa}) only by $d\to s$ and the primes on 
the hadronic matrix elements that in principle are different in these two 
decays. As
\be
V_{cs}V_{cb}^*\approx A\lambda^2, \qquad
V_{us}V_{ub}^*\approx A \lambda^4 R_b e^{i\gamma}, 
\ee
the second term in (\ref{Fgamma}) is even more important than the 
corresponding term in 
the case of $B^0_d\to \pi^+\pi^-$. Consequently $B^0_d\to K^+K^-$
taken alone does not offer a useful method for the determination of the CKM 
phases. On the other hand, with the help of the U-spin symmetry of strong 
interations, it allows roughly speaking to determine the penguin contributions
in $B^0_d\to \pi^+\pi^-$ and consequently the extraction of $\beta$ and 
$\gamma$.

Indeed, from the U-spin symmetry we have
\be
\frac{P_{\pi\pi}}{T_{\pi\pi}}=\frac{P_c-P_t}{A_T+P_u-P_t}
=\frac{P'_c-P'_t}{A'_T+P'_u-P'_t}=\frac{P_{KK}}{T_{KK}}
\equiv d e^{i\delta}
\ee
where $d$ is a real non-perturbative parameter and $\delta$ a strong phase.
Measuring $S_f$ and $C_f$ for both decays and extracting $\beta_s$ from
$B_s^0\to J/\psi \phi$, we can determine four unknows: $d$, $\delta$, $\beta$ 
and $\gamma$ subject mainly to U-spin breaking corrections. 
A recent analysis using these ideas can be found in \cite{FLISMA}.
\subsection{Constraints for \boldmath{$\gamma$} from $B\to\pi K$}
\subsubsection{Preliminaries}
The recent developments involve also the extraction of
the angle $\gamma$ from the decays $B\to \pi K$.
The modes $B^\pm\to \pi^\mp K^0$, $B^\pm\to \pi^0 K^\pm$,  
$B^0_d\to \pi^\mp K^\pm$ and $B^0_d\to \pi^0 K^0$
have been observed 
by the CLEO, BaBar  and Belle collaborations and 
should 
allow us to obtain direct information on $\gamma$ 
when the errors on branching ratios and the CP asymmetries decrease. 
The latter are still consistent with zero.
The progress on the accuracy of these measurements is 
slow but steady and they will certainly give an interesting insight 
into the flavour dynamics and QCD dynamics one day.

There has been a large theoretical activity in this field during 
the last six years.
The main issues here are the final state interactions (FSI), 
SU(3) symmetry
breaking effects and the importance of electroweak penguin
contributions. Several interesting ideas have been put forward
to extract the angle $\gamma$ in spite of large hadronic
uncertainties in $B\to \pi K$ decays 
\cite{FM,GRRO,GPAR1,GPAR3,GPAR2,NRBOUND}.

Three strategies for bounding and determining $\gamma$ have been 
proposed. The ``mixed" strategy \cite{FM} uses 
$B^0_d\to \pi^0 K^\pm$ and $B^\pm\to\pi^\pm K$. The ``charged" strategy
\cite{NRBOUND} involves $B^\pm\to\pi^0 K^\pm,~\pi^\pm K$ and
the ``neutral" strategy \cite{GPAR3} the modes 
$B_d^0\to \pi^\mp K^\pm,~\pi^0K^0$. 
General parametrizations for the 
study of the FSI, SU(3) symmetry
breaking effects and of the electroweak penguin
contributions in these channels have been presented 
in \cite{GPAR1,GPAR3,GPAR2}.
Moreover, general parametrizations by means
of Wick contractions \cite{IWICK,BSWICK} have been proposed. 
They can be used for all two-body B-decays.
These parametrizations should
turn out to be useful when the data improve.

Parallel to these efforts an important progress has been made 
by developing approaches for
the calculation of the hadronic matrix elements of local operators in 
QCD beyond the standard factorization method. These are in particular the
QCD factorization approach \cite{BBNS1}, the perturbative QCD approach
\cite{Li} and the soft-collinear effective theory \cite{SCET}.
Moreover new methods to calculate exclusive hadronic matrix
elements from QCD light-cone sum rules have been developed 
in \cite{KOD}. 
While,
in my opinion, an important progress in evaluating non-leptonic 
amplitudes has been made in these papers, the usefulness of this 
recent progress
at the quantitative level has still to be demonstrated when the
data improve.  

\subsubsection{A General Parametrization for \boldmath{$B\to\pi K$}}
In order to illustrate the complexity of the extraction of $\gamma$ from
these decays let me describe briefly the general parametrization for the 
mixed, charged 
and neutral strategies, developed in 1998 in collaboration with Robert 
Fleischer \cite{GPAR3}.

The isospin symmetry implies in each case one relation between the relevant 
amplitudes:
\be\label{mixed}
A(B^+\to \pi^+ K^0)+A(B^0_d\to \pi^-K^+)=-\left[T+P_{\rm EW}^{\rm C}\right]
\ee  
\be\label{charged}
A(B^+\to \pi^+ K^0)+\sqrt{2} A(B^+\to \pi^0 K^+)=-\left[(T+C)+P_{\rm EW}\right]
=3 A_{3/2}
\ee
\be\label{neutral}
\sqrt{2} A(B_d^0\to \pi^0 K^0)+A(B_d^0\to \pi^- K^+)=
-\left[(T+C)+P_{\rm EW}\right]
=3 A_{3/2}
\ee
where $T$ stands for tree, $C$ for colour suppressed tree, $P_{\rm EW}$ for 
electroweak penguins and $P_{\rm EW}^{\rm C}$ for colour suppressed 
electroweak penguins. $A_{3/2}$ is an isospin amplitude. In particular we 
have
\be
T+C=|T+C|e^{i\delta_{T+C}}e^{i\gamma}, 
\qquad P_{\rm EW}=-|P_{\rm EW}|e^{i\delta_{\rm EW}}
\ee
where the $\delta_i$ denote the strong interaction phases.

The QCD penguins, absent in (\ref{mixed})--(\ref{neutral}), 
enter the analysis in 
the following manner:
\be
P_{ch}\equiv A(B^+\to \pi^+ K^0)=-(1-\frac{\lambda^2}{2})\lambda^2 A 
\left[1+\varrho_{ch}e^{i\theta_{ch}} e^{i\gamma}\right]
|P^{ch}_{tc}|e^{i\delta^{ch}_{tc}}
\ee

\be
P_{n}\equiv \sqrt{2}A(B^0_d\to \pi^0 K^0)=-(1-\frac{\lambda^2}{2})\lambda^2 A 
\left[1+\varrho_{n}e^{i\theta_{n}} e^{i\gamma}\right]
|P^{n}_{tc}|e^{i\delta^{n}_{tc}}
\ee
where the terms proportional to $\varrho_{ch,n}$ parametrize $u$-penguin and 
rescattering effects and the last factors stand for the difference of $t$ and
$c$ penguins. 

The relevant parameters in the three strategies in question are
\be
r=\frac{|T|}{\sqrt{|P_{ch}|^2}}, \qquad 
q=\left|\frac{P_{\rm EW}^{\rm C}}{T}\right| e^{i\bar\omega},
\qquad \delta=\delta_T-\delta_{tc}
\ee
\be
r_{ch}=\frac{|T+C|}{\sqrt{|P_{ch}|^2}}, 
\qquad q_{ch}=\left|\frac{P_{\rm EW}}{T+C}\right|    e^{i\omega},
\qquad \delta_{ch}=\delta_{T+C}-\delta^{ch}_{tc}
\ee
\be
r_n=\frac{|T+C|}{\sqrt{|P_n|^2}}, \qquad q_n=q_{ch},
\qquad \delta_n=\delta_{T+C}-\delta^n_{tc}~.
\ee
The virtue of this general parametrization, is the universality of various 
formulae for quantities of interest, not shown here due to the lack of space.
In order to study a given strategy, the relevant parameters listed above 
have to be inserted in these formulae.

The formulae given above are not sufficiently informative for a determination
of $\gamma$. To proceed further one has to use $SU(3)$ flavour symmetry. This
allows to fix $r_{ch}$, $r_n$ and $q_{ch}=q_n$. On the other hand $r$ and $q$
are not determined by $SU(3)$ and their values have to be estimated by some 
dynamical assumptions like factorization. Consequently the mixed strategy has 
larger theoretical uncertainties than the other two strategies.

We have then, respectively, \cite{NRBOUND,rch} 
\be
q_{ch}=q_n=0.70 \left[\frac{0.37}{R_b}\right]=0.70\pm 0.08,
\ee
\be
r_{ch}=\sqrt{2}\left|\frac{V_{us}}{V_{ud}}\right|\frac{F_K}{F_\pi}
\sqrt{\frac{Br(B^\pm\to \pi^\pm\pi^0)}{Br(B^\pm\to \pi^\pm K^0)}}
=0.20\pm 0.02
\ee
with $|T+C|$ in $r_{ch}$ extracted from $B^\pm\to \pi^\pm\pi^0$.
The last number is my own estimate. Similarly 
one finds \cite{GPAR3}
\be
r_n=\left|\frac{V_{us}}{V_{ud}}\right|\frac{F_K}{F_\pi}
\sqrt{\frac{Br(B^\pm\to \pi^\pm\pi^0)}{Br(B^0_d\to \pi^0 K^0)}}
\sqrt{\frac{\tau(B^0_d)}{\tau(B^\pm)}}
=0.17\pm 0.02.
\ee

As demonstrated in a vast number of papers 
\cite{FM,GPAR1,GPAR3,GPAR2,NRBOUND}, 
these strategies imply interesting bounds on $\gamma$ 
that not necessarily agree with the values extracted from the UT analysis 
of section 3.  In particular already in 1999 combining the neutral and
charged strategies \cite{GPAR3} we have  found that the 1999 data 
on $B\to \pi K$ favour $\gamma$  in the second quadrant, which is in 
conflict with the standard analysis of the unitarity triangle that 
implied $\gamma= (62\pm 7)^\circ$. Other arguments for $\cos\gamma<0$ using
$B\to PP,~PV$ and $VV$ decays were given also in \cite{CLEO99}.
Recent  analyses of $B\to\pi K$ by various authors  find also that 
 $\gamma>90^\circ$ is favoured by the $B\to \pi K$ data. 
Most recent reviews can be found in \cite{CERNCKM}.
See also \cite{Neubert02}.

In view of sizable theoretical uncertainties in the analyses of
$B\to\pi K$ and of still significant experimental errors in the
corresponding branching ratios it is not yet clear whether the
discrepancy in question is serious. For instance \cite{CIFRMAPISI}
sizable contributions of the so-called charming penguins to the
$B\to\pi K$ amplitudes could shift $\gamma$ extracted from these
decays below $90^\circ$ but at present these contributions cannot be
calculated reliably. Similar role could be played by annihilation
contributions \cite{Li} and large non-factorizable 
SU(3) breaking effects
\cite{GPAR3}.  Also,  new physics contributions in the electroweak
penguin sector could  shift $\gamma$ to the first quadrant
\cite{GPAR3}.  It should be however emphasized that the problem with
the angle $\gamma$, if it persisted, would put into difficulties not
only the SM but also the full class of MFV models in which the lower
bound on $\Delta M_s/\Delta M_d$ implies $\gamma < 90^\circ$. 
In any case it will be exciting to follow the developments in this 
field. 

\section{\boldmath{$\kpn$} and \boldmath{$\klpn$} } 
The rare decays $\kpn$ and $\klpn$ are very promising probes of 
flavour physics within the SM and possible extensions, since they are
governed by short distance interactions. They proceed through $Z^0$-penguin
 and box diagrams. As the required hadronic matrix elements can be extracted 
from the leading semileptonic decays and other long distance contributions 
turn out to be negligible \cite{RS}, the relevant branching ratios can be 
computed
to an exceptionally high degree of precision \cite{BB,BB98,MU98}. 
The main theoretical
uncertainty in the CP conserving decay $\kpn$ originates in the value of
$\mc(\mu_c)$. It has been reduced through NLO corrections down to $\pm 7\%$ 
\cite{BB,BB98}
at the level of the branching ratio. The dominantly CP-violating decay 
$\klpn$ \cite{littenberg:89} is even cleaner as only the internal top 
contributions matter. The 
theoretical error for $Br(\klpn)$ amounts to $\pm 2\%$ and is safely 
negligible. 

\subsection{Branching Ratios}
The basic formulae for the branching ratios are given as follows
\begin{equation}\label{bkpn}
Br(\kpn)=\kappa_+\cdot\left[\left(\IM F_t \right)^2+
\left(\RE F_c +\RE F_t\right)^2 \right]~,
\end{equation}
\begin{equation}\label{bklpn}
Br(K_{\rm L}\to\pi^0\nu\bar\nu)=\kappa_{\rm L}\cdot
\left(\IM F_t  \right)^2~,
\end{equation}
where
\be 
F_c={\lambda_c\over\lambda}P_0(X), \qquad 
F_t={\lambda_t\over\lambda^5}X(x_t).
\ee
Here $\lambda_i=V^\ast_{is}V_{id}$ and
\begin{equation}\label{kapp}
\kappa_+=4.75\cdot 10^{-11}\,, \qquad
\kappa_{\rm L}=2.08\cdot 10^{-10}
\end{equation}
include isospin
breaking corrections in relating $\kpn$ and $\klpn$ to $K^+\to\pi^0e^+\nu$,  
respectively \cite{MP}.
Next
\be\label{XT}
X(x_t)=1.52\cdot\left [\frac{\mtb(\mt)}{167~GeV} \right ]^{1.15} 
\ee
represents internal top contribution and 
$P_0(X)=0.39\pm 0.06$ results from the internal charm 
contribution \cite{BB}. The numerical values in (\ref{kapp}) and for 
$P_0(X)$ differ from \cite{BB98} due to a different  value 
of $\lambda=0.224$ used here.

Imposing all existing constraints on the CKM matrix one finds 
\cite{Gino03}
\begin{equation}\label{kpnr}
Br(\kpn)=
(7.7 \pm 1.1)\cdot 10^{-11}, 
\end{equation}
\begin{equation}\label{klpnr4}
Br(\klpn)=
(2.6 \pm 0.5)\cdot 10^{-11} 
\end{equation}
where the errors come dominantly from the uncertainties in the CKM
parameters. Similar results are found in
 \cite{KENDAL}.
 The first result should be compared with the measurements of  
AGS E787 collaboration at Brookhaven \cite{Adler97} that observing
two events for this very rare decay finds
\be\label{kp01}
Br(K^+ \rightarrow \pi^+ \nu \bar{\nu})=
(15.7^{+17.5}_{-8.2})\cdot 10^{-11}~.
\end{equation}
This is a factor of 2 above the SM expectation. Even if the errors 
are substantial and the result is compatible with the SM, the branching ratio
(\ref{kp01}) implies already a non-trivial lower bound on $\vtd$ 
\cite{Adler97,AI01}. 

The present upper bound on $Br(K_{\rm L}\to \pi^0\nu\bar\nu)$ from
the KTeV  experiment at Fermilab \cite{KTeV00X} reads 
\begin{equation}\label{KLD}
Br(\klpn)<5.9 \cdot 10^{-7}\,.
\end{equation}
This is about four orders of magnitude above the SM expectation
(\ref{klpnr4}).
Moreover this bound is substantially weaker than the 
{\it model independent} bound \cite{NIR}
from isospin symmetry:
\begin{equation}
Br(\klpn) < 4.4 \cdot Br(\kpn)
\end{equation}
which through (\ref{kp01})  gives
\begin{equation}\label{B108}
Br(\klpn) < 1.6 \cdot 10^{-9} ~(90\% C.L.)
\end{equation}
\begin{figure}[hbt]
\vspace{0.10in}
\centerline{
\epsfysize=2.0in
\epsffile{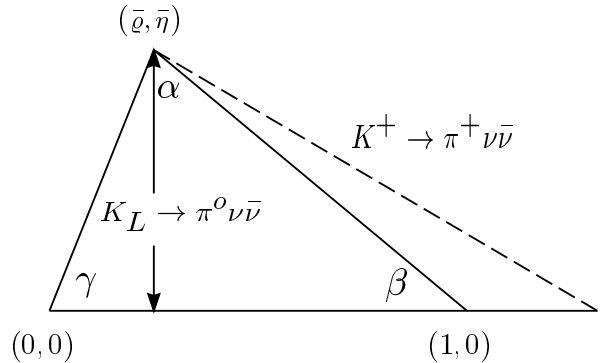}
}
\vspace{0.08in}
\caption{Unitarity triangle from $K\to\pi\nu\bar\nu$.}\label{fig:KPKL}
\end{figure}

\subsection{Unitarity Triangle and \boldmath{$\sin 2\beta$} from 
\boldmath{$K\to\pi\nu\bar\nu$}}
\label{sec:Kpnn:Triangle}
The measurement of $Br(\kpn)$ and $Br(\klpn)$ can determine the
unitarity triangle completely, (see fig.~\ref{fig:KPKL}) \cite{BBSIN}.
The explicit formulae can be found in \cite{BLO,Erice,BBSIN}. The most 
interesting in this context are very clean determinations of $\sin 2\beta$ 
and ${\rm Im}\lambda_t$ that are free not only from hadronic uncertainties 
but also parametric uncertainties like $\vcb$ and $m_c$. The determination 
of $\vtd$ is also theoretically clean but its precision depends on the 
accuracy with which $\vcb$ and $m_c$ are known. Also the scale uncertainties 
in $\vtd$ amount to $4\%$ at the NLO \cite{BB}. They should be 
significantly reduced through
a calculation of NNLO corrections to the charm contribution 
that is in progress and should be available in 2004. 

Assuming that the branching ratios will be  known to within $\pm 10\%$
we expect the following accuracy in this decade
\be
\sigma(\sin 2\beta)=\pm 0.04, \quad
\sigma({\rm Im}\lambda_t)=\pm 5\%, \quad
\sigma(|V_{td}|)= \pm 7\%~. 
\ee
The comparison with the corresponding determinations in B decays
will offer a very good test of flavour dynamics and CP violation in the
SM and a powerful tool to probe the physics
beyond it. 

\subsection{Concluding Remarks}
As the theorists were able to calculate the branching ratios for 
these decays rather precisely, the future of this field is in the 
hands of experimentalists and depends on the financial support 
that is badly needed.
The experimental outlook for these decays has been reviewed 
in \cite{LITT00,Bel}.
The future of $\kpn$ depends on the AGS E949
 and the CKM experiment at Fermilab. 
In the case of $\klpn$ these are 
the KEK E391a experiment, KOPIO
at Brookhaven (BNL E926) and
an experiment at the 50 GeV JHF in Japan that should be 
able to collect 1000 events at the end of this decade.
Both KOPIO and JHF should provide very important measurements of this
gold-plated decay. For a recent theoretical review see \cite{Gino03}.

\section{Minimal Flavour Violation Models}
\subsection{Preliminaries}
We have defined this class of models in Section 1. Here I would like just
to list four interesting properties of these models that are independent 
of particular parameters present in these models. Other relations can be
found in \cite{REL}. These are:
\begin{itemize}
\item
There exists a universal unitarity triangle (UUT) \cite{UUT} common to all 
these models and the SM that can be constructed by using measurable 
quantities that depend on the CKM parameters but are not polluted by the 
new parameters present in the extensions of the SM. 
The UUT can be constructed, for instance, by using $\sin 2\beta$ from 
$a_{\psi K_S}$ and the ratio $\Delta M_s/\Delta M_d$. 
The relevant formulae can be found in Section 3 and in
 \cite{UUT,BF01}, where also other 
quantities suitable for the determination of the UUT are discussed.
\item
\be
(\sin 2\beta)_{J/\psi K_S}=(\sin 2\beta)_{\phi K_S}
=(\sin 2\beta)_{\pi\nu\bar\nu}
\ee
\item
For given $a_{\psi K_{\rm S}}$ and $Br(\kpn)$ only two values of 
$Br(\klpn)$ are possible 
in the full class of MFV models, independently of any new parameters 
present in these models \cite{BF01}. 
Consequently, measuring $Br(\klpn)$ will 
either select one of these two possible values or rule out all MFV models.
The present experimental bound on $Br(\kpn)$ and $\sin 2\beta\le 0.80$
imply an absolute upper bound 
$Br(\klpn)<4.9 \cdot 10^{-10}~(90\%~{\rm C.L.})$ 
\cite{BF01} in the MFV models that is stronger than the bound in (\ref{B108}).
\item
There exists a correlation between $Br(B_{d,s}\to\mu\bar\mu)$ and 
$\Delta M_{d,s}$ \cite{AJB03}:
\be
\frac{Br(B_{s}\to\mu\bar\mu)}{Br(B_{d}\to\mu\bar\mu)}
=\frac{\hat B_{d}}{\hat B_{s}}
\frac{\tau( B_{s})}{\tau( B_{d})} 
\frac{\Delta M_{s}}{\Delta M_{d}} \nonumber 
\ee
that is practically free of theoretical uncertainties as 
$\hat B_{s}/\hat B_{d}=1$ up to small breaking $SU(3)$ breaking corrections.
Similar correlations between $Br(B_{d,s}\to\mu\bar\mu)$ and 
$\Delta M_{d,s}$, respectively, allow rather precise predictions for 
$Br(B_{d,s}\to\mu\bar\mu)$ within the MFV models once $\Delta M_{d,s}$ 
are known  \cite{AJB03}.
\end{itemize}
\subsection{Universal Unitarity Triangle}
The presently available quantities that do not depend on the new physics 
parameters within the MFV models and therefore can be used to determine 
the UUT are $R_t$ from $\Delta M_d/\Delta M_s$ by means of (\ref{Rt}),
$R_b$ from $\vub$ by means of (\ref{2.94})   and $\sin 2\beta$ 
extracted from the CP asymmetry in $B^0_d\to \psi K_S$. 
Using only these three quantities, we show in figure \ref{fig:figmfv}  
the allowed universal region for $(\bar\varrho,\bar\eta)$ (the larger
ellipse) in the MFV models as obtained recently in an update of 
\cite{BUPAST}. The results for various quantities of 
interest related to this UUT are collected in table \ref{mfv}.
Similar analysis has been done in \cite{AMGIISST}.

It should be stressed that any MFV model that is inconsistent with the 
broader allowed region in figure \ref{fig:figmfv} and 
the UUT column in table \ref{mfv} is ruled out. 
We observe that there is little room for MFV models that in their predictions 
for UT differ significantly from the SM. It is also clear that to 
distinguish the SM from the MFV models on the 
basis of the analysis of the UT of Section 3, will require 
considerable reduction of 
theoretical uncertainties.
\subsection{Models with Universal Extra Dimensions}
In view of the difficulty in distinguishing various MFV models on the 
basis of the standard analysis of UT from each other, it is essential to 
study other FCNC processes as rare B and K decays and radiative B decays like
$B\to X_s\gamma$ and $B\to X_s\mu^+\mu^-$. In the case of MSSM at low 
$\tan \beta$ such an analyses can be found in \cite{BRMSSM,ALIHILL}. 
Recently a very extensive analysis
of all relevant FCNC processes in a SM with one universal extra dimension 
\cite{appelquist:01} has been presented in \cite{BSW02,BPSW}. In this model 
all standard model fields can propagate in the fifth dimension and the 
FCNC processes are affected by the exchange of the Kaluza-Klein particles 
in loop diagrams.  
The most interesting 
results of \cite{BSW02,BPSW,BPSWD} are the enhancements of $Br(\kpn)$ and 
$Br(B\to X_s\mu^+\mu^-)$, strong suppressions of $Br(B\to X_s\gamma)$
and $Br(B\to X_s~{\rm gluon})$ and a significant downward shift of the 
zero $\hat s_0$ in the forward-backward asymmetry in $Br(B\to X_s\mu^+\mu^-)$.
\section{Outlook}
Let me finish these lectures with my personal expectations for 
the coming years 
with regard to the CKM matrix and FCNC processes.

\subsection{Phase 1 (2003-2007)}
In this phase the determination of the CKM matrix will be governed 
by 
\be\label{OT1}
V_{us}, \qquad \vcb,\qquad a_{\psi K_S}, \qquad \Delta M_d/\Delta M_s.
\ee
These four quantities are sufficient to determine the full CKM matrix  
and suggest  a new set of fundamental variables \cite{BUPAST}
\be\label{OT2}
V_{us}, \qquad \vcb,\qquad \beta, \qquad R_t.
\ee
The precision of this determination will 
depend on the 
accuracy with which  $a_{\psi K_S}$ and $\Delta M_d/\Delta M_s$
will be measured and the non-perturbative ratio $\xi$  calculated
by lattice and QCD sum rules methods.

An important role will also be played by 
\be
\varepsilon_K, \qquad \vub
\ee
but this will depend on the reduction of the hadronic uncertainties in
 $\hat B_K$ and in the determination of $|V_{ub}|$ 
\cite{CERNCKM}.  
Note that $\varepsilon_K$ and $\vub$ combined with (\ref{OT1}) can tell 
us whether the CP violation in the K-system is consistent with the one 
observed in the B-system. 

Very important is the clarification of the possible 
discrepancy between the measurements of the angle $\beta$ by means of
$a_{J/\psi K_S}$ and $a_{\phi K_S}$ that if confirmed would imply new 
sources of CP violation. Similarly the status of CP violation in 
$B^0_d\to\pi^+\pi^-$ and in $B\to\pi K$ decays should be clarified 
in this phase but this will depend  on the theoretical 
progress in non-perturbative methods. 
We should also be able to get some information about 
$\gamma$ not only from $B\to \pi K$ at
B factories but also  by means of
U-spin strategies in conjuction with the data from Run II at Tevatron.

During this phase we should also have new data on $Br(\kpn)$ from AGS E949 
and first data from the CKM experiment at Fermilab. 
The comparison 
of these data with the implications of  $\Delta M_d/\Delta M_s$ should be 
very interesting \cite{BB98,AI01,BSW02,FLISMA}. It would be particularly 
exciting if the central value did not decrease below the one in 
(\ref{kp01}), that is 
roughly by a factor of two higher than the SM value.
In any case these data should have a considerable impact on $\vtd$ and
the unitarity triangle.

We will also have new data on $B_s\to \mu^+\mu^-$, $B\to X_s\nu\bar\nu$, 
$B\to X_s \gamma$, $B \to X_s\mu\bar\mu$ as well as on related exclusive 
channels. All these decays are governed by the CKM element $\vts$ that is 
already well determined by the unitarity of the CKM matrix $\vts\approx \vcb$. 
Consequently I do not expect that these decays will play an important role in
the CKM fits. On the other hand being sensitive to new physics contributions 
they could give the first signals of new physics. The fact that $\vts$ is 
already reasonably well known will be helpful in this context.

\subsection{Phase 2 (2007-2009)}
With the B-factories and Tevatron entering their mature stage and LHCB,
BTeV, Atlas and CMS  beginning hopefully their operation, the quantities in 
(\ref{OT1}) should offer a very good determination of the CKM matrix. 
I expect that other decays listed in Phase 1 will become more useful in view 
of improved data and new theoretical ideas.
The most important new developments to be expected in this phase will be clean 
measurements of the angle $\gamma$ at LHCB and BTeV in decays 
$B_s\to D^+_s K^{-}$ and $\bar B_s\to D^-_s K^{+}$
 and an 
improved measurement of $Br(\kpn)$ by the CKM collaboration at Fermilab. 
Also other strategies discussed in Section 5.4 and possible 
measurements of rare B-decays sensitive to both $\vts$ and $\vtd$ should play
an important role. This phase 
should provide (in case the phase 1 did not do it) definite answer 
whether MFV is sufficient to describe the 
data or whether new flavour violating interactions are required.

At the end of this phase we should also have much more improved knowledge 
about $\klpn$ from KOPIO at Brookhaven and JHF in Japan.

However, the most interesting scenario would be the discovery of
supersymmetry at LHC which could considerably reduce the uncertainty in 
the supersymmetric parameters necessary for the study of FCNC processes.

\subsection{Phase 3 (2009-2013)}
Here precise measurements of $Br(\klpn)$  from KOPIO and
JHF will be among 
the highlights. In addition the branching ratios for most of the decays
studied in phases 1 and 2 will be known with much higher precision. This will
allow not only a precision test of SM but also to 
identify the patterns of new physics contributions that I personnally expect 
should show up at this level of accuracy. The combination of these studies 
with the results from LHC that should signal some direct signs of new physics
should allow a convincing identification of this new physics.

No doubt the next ten years should be very exciting but the real progress 
will require extreme joined efforts by theorists and experimentalists.

{\bf Acknowledgements}

I would like to thank the organizers for inviting me to such a 
wonderful winter school  and most enjoyable atmosphere. 
I would also like to thank Robert Fleischer, Matthias Jamin,  
 Stefan Recksiegel and Achille Stocchi for discussions.
The work presented here has been supported in part by the German 
Bundesministerium f\"ur
Bildung und Forschung under the contract 05HT1WOA3 and the 
DFG Project Bu. 706/1-2.

%


\begin{thebibliography}{80}
\bibitem{WARN} 
The numerical value on the r.h.s of (\ref{100}) being proportional to
$\lambda^{-10}$ is very sensitive to $\lambda$ but as $A\propto \lambda^{-2}$
and $P_c(\varepsilon)\propto \lambda^{-4}$ the constraint (\ref{100}) depends
only quadratically on $\lambda$.
\end{thebibliography}
\end{document}